\documentclass{aa}

\usepackage{txfonts}
\usepackage{graphics}
\usepackage{epsfig}
\usepackage{pdflscape}

\begin{document}

\title{The VMC survey - XVII. The proper motion of the Small Magellanic Cloud and of the Milky Way globular cluster 47 Tucanae \thanks{Based on observations made with VISTA at the Paranal Observatory under program ID 179.B-2003.}}

\author{Maria-Rosa L. Cioni\inst{1, 2, 3}
	\and Kenji Bekki\inst{4}
	\and L\'{e}o Girardi\inst{5}
	\and Richard de Grijs\inst{6,7}
	\and Mike J. Irwin\inst{8}
	\and Valentin D. Ivanov\inst{9,10}
	\and Marcella Marconi\inst{11}
	\and Joana M. Oliveira\inst{12}
	\and Andr\'{e}s E. Piatti\inst{13,14}	
	\and Vincenzo Ripepi\inst{11}
	\and Jacco Th. van Loon\inst{12}
}

\offprints{mcioni@aip.de}

\institute{
	Universit\"{a}t Potsdam, Institut f\"{u}r Physik und Astronomie, Karl-Liebknecht-Str. 24/25, D-14476 Potsdam, Germany
	\and Leibnitz-Institut f\"{u}r Astrophysik Potsdam, An der Sternwarte 16, D-14482 Potsdam, Germany
	\and University of Hertfordshire, Physics Astronomy and Mathematics, College Lane, Hatfield AL10 9AB, United Kingdom
	\and ICRAR, M468, University of Western Australia, 35 Stirling Hwy, Crawley 6009, Western Australia, Australia
	\and INAF, Osservatorio Astronomico di Padova, vicolo dell'Osservatorio 5, I-35122 Padova, Italy
	\and Kavli Intitute for Astronomy \& Astrophysics, Peking University, Yi He Yuan Lu 5, Hai Dian District, Beijing 100871, China
	\and Department of Astronomy, Peking University, Yi He Yuan Lu 5, Hai Dian District, Beijing 100871, China
	\and Institute of Astronomy, University of Cambridge, Madingley, Road, Cambridge, CB3 0HA, UK
	\and European Southern Observatory, Karl-Schwarzschild-Str. 2 , D-85748 Garching bei M\"{u}nchen, Germany
	\and European Southern Observatory, Ave. Alonso de C\'ordova 3107, Vitacura, Santiago, Chile
	\and INAF, Osservatorio Astronomico di Capodimonte, via Moiariello 16, Napoli, I-80131, Italy
	\and Lennard-Jones Laboratories, School of Physical and Geographical Sciences, Keele University, ST5 5BG, United Kingdom
	\and Observatorio Astron\'{o}mico, Universidad Nacional de C\'{o}rdoba, Laprida 854, 5000, C\'{o}rdoba, Argentina
	\and Consejo Nacional de Investigaciones Cient\'{i}ficas y T\'{e}cnicas, Av. Rivadavia 1917, C1033AAJ, Buenos Aires, Argentina
	}

\date{Received  / Accepted }

\titlerunning{47 Tuc proper motion}

\authorrunning{Cioni et al.}

\abstract
{}
{In this study we use multi-epoch near-infrared observations from the VISTA survey of the Magellanic Cloud system (VMC) to measure the proper motion of different stellar populations in a tile of $1.5$ deg$^2$ in size in the direction of the Galactic globular cluster 47 Tuc. We obtain the proper motion of the cluster itself, of the Small Magellanic Cloud (SMC), and of the field Milky Way stars.}
{Stars of the three main stellar components are selected from their spatial distribution and their distribution in colour-magnitude diagrams. Their average coordinate displacement is computed from the difference between multiple $K_\mathrm{s}$-band observations for stars as faint as $K_\mathrm{s}=19$ mag. Proper motions are derived from the slope of the best-fitting line among $10$ VMC epochs over a time baseline of $\sim1$ yr. Background galaxies are used to calibrate the absolute astrometric reference frame.}
{The resulting absolute proper motion of 47 Tuc is $(\mu_\alpha\cos(\delta)$, $\mu_\delta)=(+7.26\pm0.03$, $-1.25\pm0.03)$ mas yr$^{-1}$. This measurement refers to about $35000$ sources distributed between $10^\prime$ and $60^\prime$ from the cluster centre. For the SMC we obtain $(\mu_\alpha\cos(\delta)$, $\mu_\delta)=(+1.16\pm0.07$, $-0.81\pm0.07)$ mas yr$^{-1}$ from about $5250$ red clump and red giant branch stars. The absolute proper motion of the Milky Way population in the line-of-sight ($l =305.9$, $b =-44.9$) of this VISTA tile is $(\mu_\alpha\cos(\delta)$, $\mu_\delta)=(+10.22\pm0.14$, $-1.27\pm0.12)$ mas yr$^{-1}$ and results from about $4000$ sources. Systematic uncertainties associated to the astrometric reference system are $0.18$ mas yr$^{-1}$. Thanks to the proper motion we detect 47 Tuc stars beyond its tidal radius.} 
%A proper motions difference of $\sim1$ mas yr$^{-1}$ between intermediate-age and old SMC stars may be related to the complex dynamical history of the galaxy.}
{}


\maketitle

\section{Introduction}
Measuring accurate proper motions of stellar populations is a current major astronomical challenge. This has become possible using state-of-the-art instrumentation that provides high accuracy and stability in time. Observations with the {\it Hubble Space Telescope} (HST) have derived the proper motions of Milky Way (MW) satellite galaxies \citep[e.g.][]{2013ApJ...768..139S, 2013ApJ...779...81M}, including the Magellanic Clouds \citep[e.g.][]{2013ApJ...764..161K}, of the Sagittarius stream \citep{2015ApJ...803...56S}, and of globular clusters \citep[e.g.][]{2014ApJ...797..115B}. The HST will continue to provide proper motion measurements for more systems while in the near future, a major contribution to proper motion studies is expected from the European Space Agency mission Gaia. 

In this context, the main advantage of proper motion studies using ground-based telescopes is to provide measurements for considerably larger samples of stars compared to those targeted with space telescopes (except for Gaia which is an all-sky survey). These studies will allow to investigate the internal kinematics of galaxies as a function of stellar population age and to relate their geometry to their dynamical history. The Magellanic Clouds represent an ideal target for this type of investigation because they are rich in stars of different ages, span a large extent on the sky, and their morphology has been severely influenced by their dynamical interaction. The Small Magellanic Cloud (SMC), in particular, has an intriguing structure with a large extent along the line-of-sight \citep[e.g.][and references therein]{2012ApJ...744..128S, 2013ApJ...779..145N, 2015AJ....149..179D} and a morphology that clearly depends on stellar population age \citep[e.g.][]{2000A&A...358L...9C, 2000ApJ...534L..53Z, 2014MNRAS.442.1663D, 2015A&A...573A.135S, 2015MNRAS.449.2768D, 2015MNRAS.449..639R}. The SMC has been recognised as a complex object to study and this explains the fewer number of studies compared to its companion galaxy, the Large Magellanic Cloud (LMC). SMC material (stars and gas) is the dominant component of the Bridge \citep[e.g.][]{1984IAUS..108..125M, 1985Natur.318..160I, 2013A&A...551A..78B} while both SMC and LMC gas has contributed to the formation of the Stream \citep[e.g.][]{2008ApJ...679..432N, 2013ApJ...772..111R}. 
%The Bridge and Stream associated with the system result both from dynamical interaction processes. 
The intermediate-age SMC stars have probably experienced substantial stripping \citep{2014MNRAS.442.1663D} and the last LMC-SMC interaction may have also brought SMC stars into the LMC \citep{2011ApJ...737...29O}. Therefore, we expect the internal motion of SMC stars to reflect the dynamical history of the galaxy. Young stars are perhaps still closely associated with structures strongly linked to their formation, while old stars, that dominate the outer regions of the galaxy, may have suffered more from tidal events. We can find out about this by studying the distribution and motion of different stellar populations across the system and produce models of the structure, kinematics and dynamics to interpret the results.

In this paper we present the proper motions of stars obtained from the analysis of multi-epoch near-infrared observations from the VISTA survey of the Magellanic Clouds system \citep[VMC;][]{2011A&A...527A.116C} in a tile towards the Galactic globular cluster 47 Tuc. This paper is part of a comprehensive investigation of the proper motion of the Magellanic Clouds using the VMC data. The main motivation of this study is to understand the past and future orbit of the Magellanic Clouds around the MW. In a first pilot study, the proper motions of different types of LMC stars, including variable stars, (within a tile towards the south ecliptic pole) were measured \citep{2014A&A...562A..32C}.

The structure of the paper is as follows. Section \ref{data} describes the VMC data used for this study. In Sect.~\ref{populations} different stellar populations are selected from their distribution in colour and magnitude as well as distance from the cluster centre. In Sect.~\ref{reference} the absolute astrometric reference frame is derived while proper motions are calculated in Sect.~\ref{pmsec}. The results of this investigation are reported in Sect.~\ref{results} while Sect.~\ref{conclusions} provides our conclusions. Finally, the Appendix lists the selection criteria adopted to distinguish the different stellar populations and shows figures of the proper motions as a function of colour.

\section{Observations and Data Selection}
\label{data}

\begin{table*}
\caption{Information of the $K_\mathrm{s}$ band observations of the VMC tile SMC $5\_2$ containing the 47 Tuc star cluster.}
\scriptsize
\label{tab52}
\[
\begin{array}{cccccccccccc}
\hline \hline
\noalign{\smallskip}
\mathrm{Tile}  & \alpha & \delta & \mathrm{\phi} & \mathrm{N} & \mathrm{First\,Epoch} & \mathrm{Last\,Epoch} & \mathrm{Time\,Interval} & \mathrm{Airmass} & \mathrm{FWHM} & e & \mathrm{Sensitivity}\\
 & \mathrm{(h:m:s)} & (^\circ \mathrm{:}^\prime\mathrm{:}^{\prime\prime}) & \mathrm{(deg)} & \mathrm{(epochs)} &  & & \mathrm{(days)} & & & & \mathrm{(mag)}\\
\noalign{\smallskip}
\hline
\noalign{\smallskip}
\mathrm{SMC\,5\_2} & 00\mathrm{:}26\mathrm{:}41.688 & -71\mathrm{:}56\mathrm{:}35.880 &   -5.5717 & 11 & 16/11/2011 & 10/11/2012 & 360 & 1.52\pm0.05 & 0.89\pm0.07 & 0.07\pm0.01 & 19.83^1\\
\noalign{\smallskip}
\hline
\hline
\end{array}
\]
$^1$ All sources with $\sigma_{K_\mathrm{s}}<0.1$ mag.
\end{table*}

\begin{figure*}
\resizebox{\hsize}{!}{\includegraphics{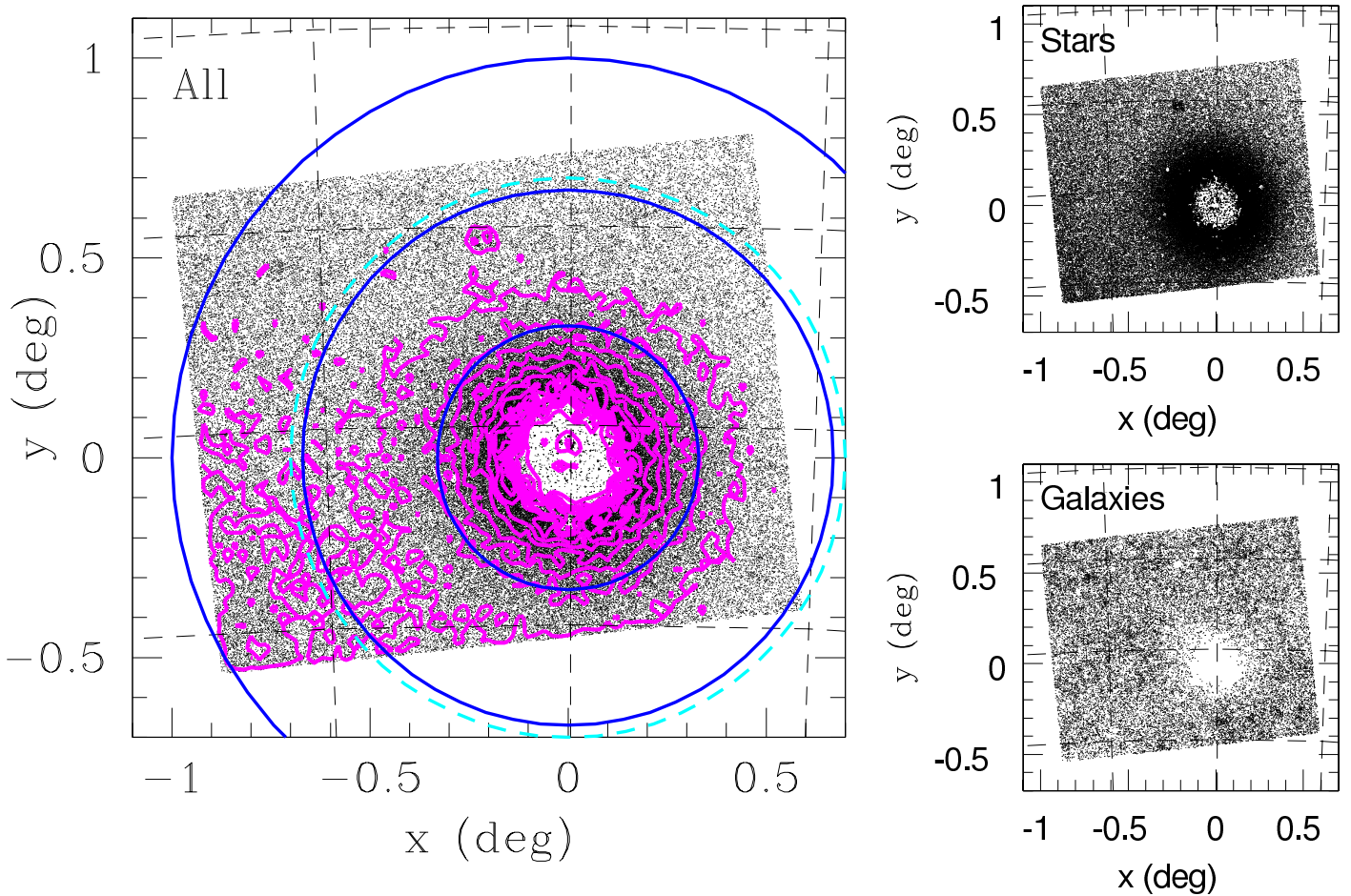}}
\caption{Distribution of all $JK_\mathrm{s}$-VISTA sources with photometric uncertainties $<0.1$ mag (left), with a star-like morphology (top right) or a galaxy-like morphology and $(J-K_\mathrm{s})>1$ (bottom right), over tile SMC $5\_2$. Note that the left diagram shows unique sources while the right diagrams refers to independent proper motion measurements where each source is present at least twice due to the detector overlaps. 
The star cluster 47 Tuc occupies about $1/4$ of the area while the SMC star cluster NGC 121 is located north of it. The centre of the plots at $(0,0)$ corresponds to the centre of 47 Tuc at $(\alpha_0$, $\delta_0)=6^\circ.023625$, $-72^\circ.081283$ (J2000) and dashed lines indicate lines of constant right ascensions and declinations. Contour levels indicate stellar population densities from $20$ to $500$ sources in steps of $20$. A dashed circle at $42^\prime$ from the centre indicates the tidal radius of the cluster while other circles correspond to distances, from the centre, of $20^\prime$, $40^\prime$, and $60^\prime$. The SMC stellar population increases to the south east of it. North is at the top and East to the left.}
\label{map}
\end{figure*}

The VMC data analysed in this study refer to observations acquired with the Visible and Infrared Survey Telescope for Astronomy \citep[VISTA;][]{2010Msngr.139....2E}. The data were reduced onto the VISTA photometric system, which is close to the Vegamag system, with the VISTA Data Flow System pipeline v1.3 \citep[VDFS;][]{2004SPIE.5493..411I} and extracted from the VISTA Science Archive \citep[VSA;][]{2012A&A...548A.119C}. The VMC survey strategy involves repeated observations of $110$ tiles (rectangular fields) across the Magellanic system, where one tile covers uniformly an area of $\sim1.5$ deg$^2$ in a given wave band with $3$ epochs at $Y$ and $J$, and $12$ epochs at $K_\mathrm{s}$ spread over a time range of a year or longer. Eleven of the $K_\mathrm{s}$ epochs refer to the monitoring campaign and each corresponds to exposure times of $750$ s (deep) while the twelfth corresponds to two observations with half the exposure time (shallow) that are not necessarily obtained during the same night. 
%Additional observations may be available, for example to recover observations redone when the original ones did not meet the sky quality requirements (transparency, seeing, and airmass). 
VMC data are acquired in service mode, hence under homogeneous sky conditions. 

Details on the observing strategy, the data reduction, and the tile pattern are given in \citet{2011A&A...527A.116C}. Here, we summarise only the relevant aspects for the proper motion analysis. Each VMC tile is the result of stacking $6$ individual pawprint observations, each containing $16$ detector images obtained from the combination of multiple exposures at five different jitter positions \citep{2004SPIE.5493..411I}. This process uses a dribbling technique to distribute the information of overlapping pixels onto the grid of the tile \citep{2009UKIRT.25.15}. In a tile, each source is detected from two to six times due to the pawprint overlaps in the mosaic reconstruction, except for two wings in the Y direction with single detections. Overlapping neighbouring tiles provide additional detections in those areas. The astrometric distortions in VISTA images are corrected based on 2MASS data and so the residuals are dominated by 2MASS errors. The astrometry of tiles suffers from a $10-20$ mas systematic pattern due to residual World Coordinate System (WCS) errors from the pawprints and a residual radial distortion of up to $\pm100$ mas across the field due to an inconsistent use of the zenithal-polynomial projection (for tiles observed prior to August 2012). 

Here we focus on one region, tile SMC $5\_2$ located in the northwestern part of the SMC and containing the Galactic globular cluster 47 Tucanae (47 Tuc). 
Table \ref{tab52} provides information on tile SMC $5\_2$: centre, orientation, number of $K_\mathrm{s}$ deep epochs available, observing date of first and last epoch, time interval, average airmass of observations, FWHM, source ellipticity, and $K_\mathrm{s}$ sensitivity resulting from the combined deep tile image.

Sources were first selected from the {\it vmcsource} merged catalogue containing sources extracted from {\it tiledeepstacks} in the $Y$, $J$, and $K_\mathrm{s}$ bands, respectively. These are deep tile images resulting from the combination of individual tile images taken at different observing epochs. Only objects detected in the $J$ and $K_\mathrm{s}$ bands with photometric uncertainties $<0.1$ mag, regardless of their $Y$-band detection, were considered. There are $203940$ objects that satisfy these criteria of which $62\%$ are most likely stars and $16\%$ galaxies, the remaining $22\%$ are possibly stars, galaxies, or extended objects resulting from blended sources. 
Figure \ref{map} shows the distribution of VMC stars and galaxies across tile SMC $5\_2$ projected with respect to the centre of 47 Tuc. The main cluster area is fully contained within the tile and stars in its outer regions are clearly detected by the VDFS pipeline, while crowding prevents the detection of stars in the inner $10\times10$ arcmin$^2$. 
North of 47 Tuc is the oldest SMC star cluster NGC 121 \citep[e.g.][]{2008AJ....135.1106G} while contour levels increasing to the southeast of 47 Tuc indicate the direction to the SMC main body. Subsequently, the same $203940$ sources were extracted from the multiple individual pawprint images in the deep $K_\mathrm{s}$ observations and only those with photometric uncertainties $<0.1$ mag in $K_\mathrm{s}$ and detection quality flags (ppErrBits) $<256$ at each epoch were retained. By working with pawprints we avoid the systematic uncertainty of $\pm100$ mas mentioned above. The choice of the ppErrBits selects only sources with minor quality issues, de-blended sources and sources with at least one bad pixel and/or a low average confidence level in the default aperture. Pawprints also avoid extra systematic errors accruing during the tiling resampling which relies on the independent pawprint WCS solutions.

\section{Stellar Populations}
\label{populations}

\begin{figure}
\resizebox{\hsize}{!}{\includegraphics{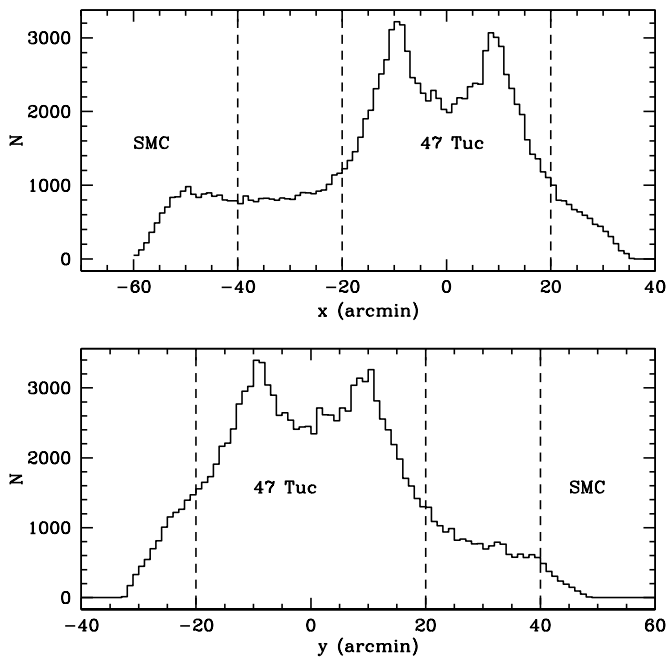}}
\caption{Distribution of VMC sources over tile SMC $5\_2$ as a function of distance along the x-axis (top) and the y-axis (bottom) from the centre of 47 Tuc. Bins are $1^\prime$ wide and vertical lines indicate areas dominated by 47 Tuc and SMC stars, respectively. Foreground and background sources have not been subtracted, but their distribution is more or less constant at all radii.}
\label{histo}
\end{figure}

The aims of this paper are to derive the proper motion of the 47 Tuc star cluster and to disentangle the influence of SMC and MW stars, using background galaxies to express the results in absolute terms. 
We have divided the observed area into three parts as a function of radius from the 47 Tuc centre (Fig.~\ref{map} left). Figure \ref{histo} shows that stars of 47 Tuc dominate the distribution at distances $\rho<20^\prime$, the region $20^\prime<\rho<40^\prime$ is characterised by a mixed population of 47 Tuc and SMC stars, while for $\rho>40^\prime$ there is a majority of SMC stars. The choice of these limits is arbitrary and meant to provide an initial characterisation of the stellar populations without influencing the subsequent results.
Foreground MW stars and background galaxies are homogeneously distributed across the three areas. The maximum cluster radius fully covered within the tile is at $\sim27^\prime$ and beyond it an increasing number of stars will be located outside our surveyed area. The tidal radius of 47 Tuc ($56$ pc) at a distance modulus of $(m-M)_0=13.40$ mag or a distance of $4.6$ kpc corresponds to $\sim42^\prime$ \citep{1996AJ....112.1487H, 2012MNRAS.423.2845L, 2014ApJ...790...35L}. 

\begin{figure*}
\resizebox{16cm}{16cm}{\includegraphics{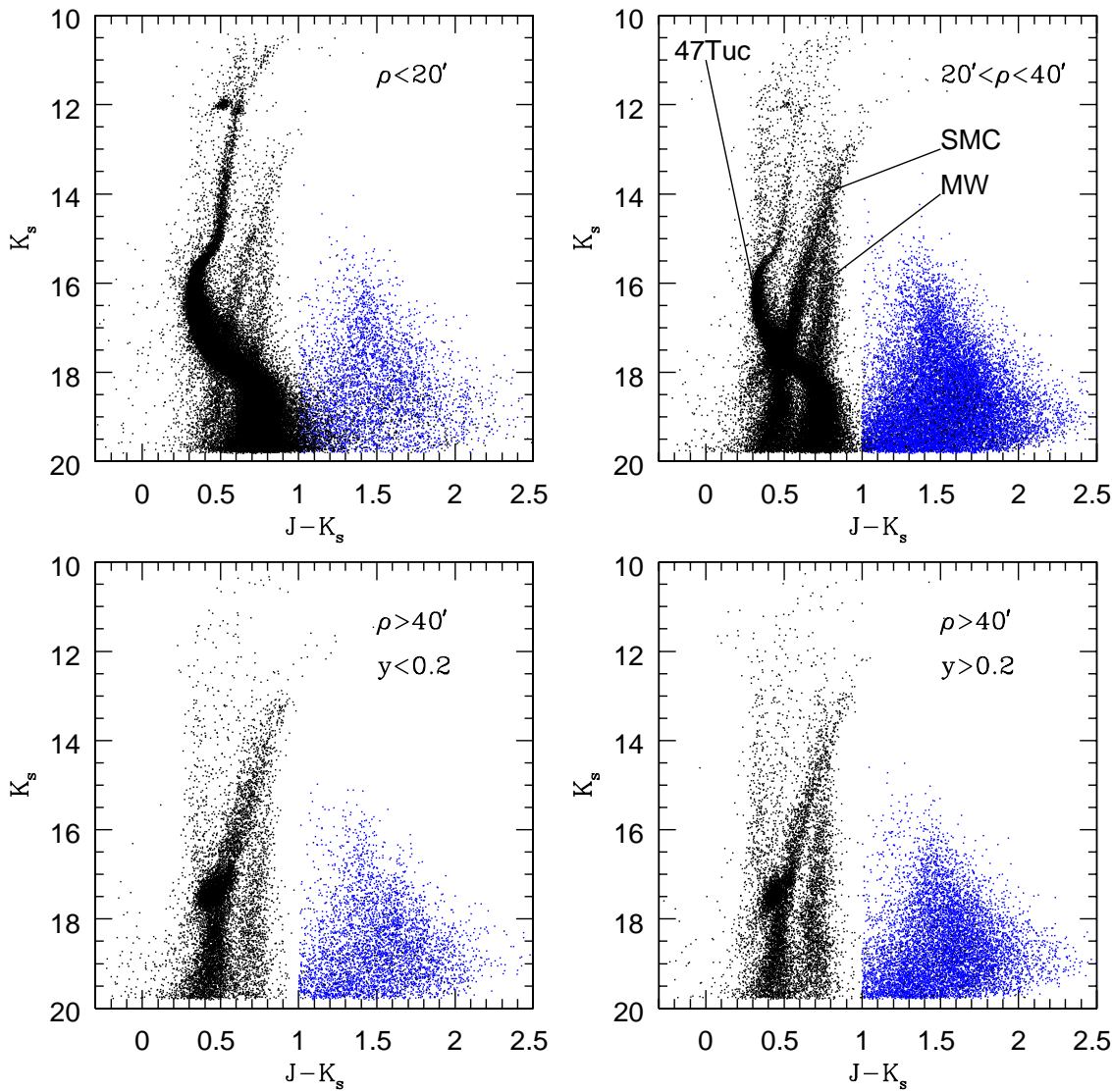}}
\caption{CMDs of VMC sources in tile SMC $5\_2$ at different radii from the 47 Tuc centre: within $20^\prime$ (top-left), between $20^\prime$ and $40^\prime$ (top-right), and further than $40^\prime$ (bottom). For the latter we distinguish between stars closest to the SMC main body (bottom-left) and in the remaining outer region of the tile (bottom-right). The three main stellar populations (47 Tuc, SMC and MW and) are indicated in top-right panel. Stars are shown in black and galaxies, with $(J-K_\mathrm{s})>1$ mag and $K_\mathrm{s}>12.5$ mag, in blue.}
\label{cmds}
\end{figure*}

The colour-magnitude diagrams (CMDs) corresponding to the three regions are shown in Fig.~\ref{cmds}. The innermost region (top-left diagram) shows a well defined locus of 47 Tuc stars. They trace the main sequence down to about two magnitudes below the turn-off ($K_\mathrm{s}\sim15.5$ mag), the sub-giant branch (SGB), an extended red giant branch (RGB), and a red clump at $K_\mathrm{s}\sim12$ mag \citep[e.g.][]{2004A&A...424..199B}. The 47Tuc CMD, based on VMC data, was first shown by \citet{2014ApJ...790...35L} with particular focus on the SGB and RGB.
Contamination by SMC and MW stars, which is limited to fairly well-defined sequences in the CMD space, is small and their colours and magnitudes are better seen at larger distances from the cluster centre (bottom-left diagram). Here, the SMC has a prominent red clump and an extended RGB; the tip of the RGB is at $K_\mathrm{s}\sim13$ mag \citep[e.g.][]{2000A&A...359..601C}. Foreground MW stars occupy a vertical sequence centred at $(J-K_\mathrm{s})=0.7$ mag, but some are also present in a bluer sequence at  $K_\mathrm{s}<16$ mag and $(J-K_\mathrm{s})$ between $0.3$ and $\sim0.5$ mag. There is no clear difference between the CMD of SMC stars in the densest region (bottom-left diagram) and those distributed in the rest of the tile (bottom-right diagram). At intermediate regions (top-right diagram) all sequences, 47 Tuc, SMC, and that of MW stars, are well populated. There are regions of the CMD where the three sequences are distinct but there are also regions of overlap, especially between the red clump of the SMC and the main sequence of 47 Tuc or between the lower part of the red MW sequence and the lower part of the 47 Tuc main sequence. The CMDs of all regions also show the distribution of background galaxies. These sources were selected from objects morphologically classified as non-stellar and after imposing a $(J-K_\mathrm{s})>1$ mag cut, to remove stellar blends. The most probable morphological classification of objects is defined from the analysis of the curve-of-growth of their flux\footnote{http://casu.ast.cam.ac.uk/surveys-projects/vista/technical/data-processing/design.pdf/view}.  The colour selection causes a sharp transition in the CMD (e.g. Fig.~\ref{cmds} top-right). These background galaxies describe a conical structure peaking at $(J-K_\mathrm{s})\sim1.4$ mag.

\begin{figure}
\resizebox{\hsize}{!}{\includegraphics{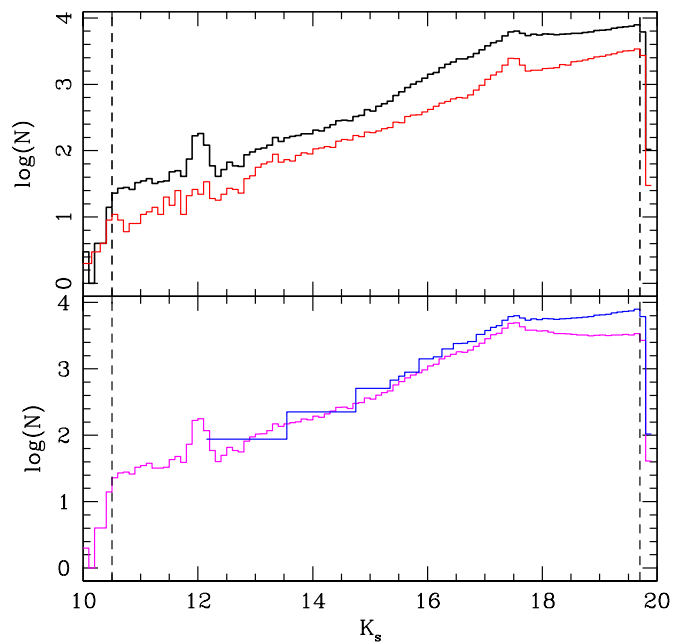}}
\caption{Distribution of VMC sources in tile SMC $5\_2$ as a function of magnitude. Bins are $0.1$ mag wide. The top panel shows the distribution of all sources (upper histogram) and of the sources excluding the main area occupied by 47 Tuc ($\rho>27^\prime$; lower histogram). The bottom panel shows the distribution for sources classified as stars (magenta) and galaxies (blue), with $(J-K_\mathrm{s})>1$ mag and $K_\mathrm{s}>12.5$ mag, by the VDFS pipeline. Vertical lines delimit the range of highest completeness.}
\label{lumf}
\end{figure}

In Fig. \ref{lumf} we plot the magnitude distribution of all sources with and without the contribution of the main 47 Tuc population (top panel), of stars and galaxies (bottom panel). The peak at $K_\mathrm{s}=12$ mag corresonds to the red clump of 47 Tuc while the red clump of the SMC is at $K_\mathrm{s}\sim17.5$ mag and the discontinuity at $K_\mathrm{s}\sim13$ mag corresponds to the tip of the RGB of the SMC. At the bright end the distribution is limited by saturation and at the faint end by the selection of sources with photometric errors $<0.1$ mag. The latter is responsible for a lack of sources at magnitudes fainter than the SMC red clump and a comparison between the histograms in Fig.~\ref{lumf} (top panel) shows that this effect is more prominent in the more crowded regions of the tile. 
In this study we use all sources in the range $11.8<K_\mathrm{s}<19.3$ mag (Fig.~\ref{pmcmd}) which are comprised in the region of highest completeness. We do not correct for foreground reddening nor for differential reddening within the SMC and 47  Tuc because we expect that observations in the $K_\mathrm{s}$ band are less sensitive to dust absorption. The foreground absorption given in the NASA/IPAC Extragalactic Database amounts to $A_V=0.101$ mag, $A_J=0.026$ mag and $A_{K_\mathrm{s}}=0.011$ mag assuming the \cite{1999PASP..111...63F} reddening law. The total absorptions in the external SMC region and the total reddening in 47 Tuc are low, $A_V=0.35$ mag  \citep{2015MNRAS.449..639R} and $E(B-V)=0.04$ mag \citep{1996AJ....112.1487H} for the SMC and 47 Tuc, respectively.

\section{Astrometric Reference Frame}
\label{reference}

We used sources classified as non-stellar and with $(J-K_\mathrm{s})>1$ to define the reference frame for the calculation of the proper motion. These are mostly galaxies that represent a non-moving system. First, these galaxies were selected from the {\it vmcsource} merged catalogue of sources, detected in the $J$ and $K_\mathrm{s}$ bands with photometric uncertainties $<0.1$ mag, according to the colour and morphology criteria. Then, six catalogues were extracted from the VSA containing each the $K_\mathrm{s}$ band counterpart of galaxies at each of the six pawprint exposures making up each tile. Subsequently, each catalogue was divided into $16$ separate catalogues, one per detector. Finally, each detector catalogue was divided into a further $11$ different catalogues, one per deep epoch of observation; shallow epochs and epochs obtained at sky conditions that did not meet the general VMC criteria were excluded. Only galaxies detected at all of the $11$ epochs and with detection flags $<256$ were selected. This flag restriction comprises of objects with a good detection quality and excludes those located in the underexposed outer region of detectors resulting from a jittering sequence. We note that the way observations are taken ensures that each of the six pawprints of the various epoch tiles are centred at (almost) the same location with respect to the optical axis of the telescope and also at the same sky position angle, otherwise the differential radial distortion from the Zenithal Polynomial projection would become an issue.

\begin{table}
\caption{Example of coefficients of the {\it geomap} transformation.}
\scriptsize
\label{fit}
\[
\begin{array}{cccc}
\hline \hline
\noalign{\smallskip}
\multicolumn{2}{c}{\mathrm{Shift}}  & \multicolumn{2}{c}{\mathrm{Rotation\,Angle}} \\ 
\multicolumn{2}{c}{\mathrm{(pix)}} & \multicolumn{2}{c}{\mathrm{(deg)}} \\ 
 x & y & x & y \\ 
\noalign{\smallskip}
\hline
\noalign{\smallskip}
-2.177\pm1.154 & 0.367\pm0.650 & 1.00004\pm0.00007 & 1.0000\pm0.0001 \\ 
\noalign{\smallskip}
\hline
\hline
\noalign{\smallskip}
\multicolumn{2}{c}{\mathrm{Magnification}} & \multicolumn{2}{c}{\mathrm{rms}}\\
\multicolumn{2}{c}{} & \multicolumn{2}{c}{\mathrm{(pix)}}\\
x & y & x &  y \\
\noalign{\smallskip}
\hline
\noalign{\smallskip}
0.009\pm0.005 & 0.010\pm0.006 & 0.196\pm0.047 & 0.198\pm0.058\\
\noalign{\smallskip}
\hline
\hline
\end{array}
\]
\end{table}

The IRAF\footnote{IRAF is distributed by the National Optical Astronomy Observatory which is operated by the Association of Universities for Research in Astronomy (AURA) under a cooperative agreement with the US National Science Foundation} tasks {\it xyxymatch}, {\it geomap}, and {\it geoxytran} were used to match the $11$ catalogues, separately per detector and pawprint, to map the reference coordinate system of $10$ individual epochs to the system of the first epoch, and finally to modify the coordinates of the galaxies at the $10$ epochs accordingly to the derived transformations. The result was $11$ catalogues for each detector ($16$) of each pawprint ($6$) of galaxies where their coordinates were given with respect to the same system, viz. that defined by the first epoch.  
A tolerance of $9$ pix and a separation of $0$ pix, to define the distance between two matching galaxies, were used. 
There were $200-400$ galaxies per detector, except for the detector dominated by the 47 Tuc cluster core (detector \#$7$) which contained only about a dozen of candidate galaxies. A linear  transformation function was fitted to the pairs of coordinates defined by the first epoch and any of the subsequent epochs; the process was iterated once rejecting galaxies exceeding the $3\sigma$ level. The function has the form: 
\begin{equation}
x^\prime = a +  bx + cy
\end{equation}
\begin{equation}
y^\prime = d + ex + fy
\end{equation}
where $a,b,c,d,e,f$ are coefficients defining the shifts, rotation angles, and scale variations of the linear transformations necessary to express the coordinates ($x,y$) of galaxies at a given epoch in the system of the first epoch ($x^\prime, y^\prime$); Table \ref{fit} shows as an example the values for pawprint \#1 and detector \#1. The {\it rms} of the residuals of the transformations was always $<0.2$ pix except for one pair (TK1-TK5) for which it was $\ge 0.3$ pix. The sensitivity of the TK5 epoch was at least $0.4$ mag shallower than that of the other epochs. Similarly, the airmass was $\sim 0.1$ larger. Both these factors indicated that epoch TK5 was of poor quality. There were two other epochs (TK4 and TK8) for which the {\it rms} was in some cases $>0.2$ pix, but their sensitivity was just $\sim 0.05$ mag lower than the sensitivity of the other epochs, and the airmass was comparable. The FWHM across the TK4 and TK8 tiles had values that were smaller and larger, respectively, than the FWHM of other tiles. The single iteration performed resulted in excluding from the position fitting procedure a handful of sources. This improved the fit of pairs involving a tile/epoch of poor quality and did not influence the other pairs. 

The lists of stars in the three areas defined in Sect. \ref{populations} as a function of radius from the cluster centre were also divided into $6\times16\times11=1056$ separate catalogues, per pawprint, detector, and epoch. The coordinates of stars in each catalogue were then adjusted to the reference system of the first epoch using the transformations derived from galaxies in the same regions. Epoch TK5 was excluded from the analysis. We emphasise that at the detector level within each pawprint, since each epoch is pointing close to the same sky location and at the same sky position angle, differential non-linear field distortions over the detector are minimal and furthermore affect both stars and galaxies equally.  This allows a more stable linear mapping to be used between the epochs taking advantage of the rigid x-y coordinate plane defined by the detectors.

\section{Proper Motion Calculations}
\label{pmsec}

\begin{figure}
\resizebox{8cm}{8cm}{\includegraphics{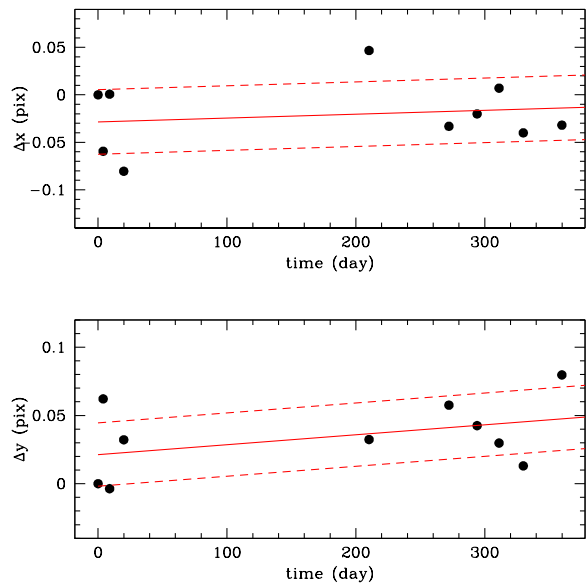}}
\caption{Coordinate differences with respect to first epochs as a function of time for a random SMC source. Contiguous lines represent the linear least squares fits through them and their slope corresponds to the proper motion in pix day$^{-1}$ in the $x$ and $y$ directions. 
%Only ten points per direction are shown because one epoch of poor quality was excluded from the calculations. 
The rms of the fits is on average $0.12$ pix and is indicated by dashed lines.}
\label{pmex}
\end{figure}

The proper motion of each source, expressed in pix day$^{-1}$, was obtained from the slope of a linear least squares fit through the coordinate differences between the first and each of  the other epochs as a function of time, separately for the two components ($x$ and $y$) and for pawprints and detectors; epoch TK5 was excluded because of poor quality (Sect.~\ref{reference}). In performing the fit both slope and intercept were allowed to vary, an example is shown in Fig.~\ref{pmex}. The proper motions were translated to mas yr$^{-1}$ by multiplying by $365.25$, to account for the number of days in a year, and by using the CD matrix parameters, that account for the rotation of the field in the sky. These parameters were taken from the header of the individual detector images for each pawprint at the reference epoch because all other epochs are already adjusted to it. The quantities $\xi$ and $\eta$ which represent the standard coordinates are defined as follows:
\begin {equation}
\label{e1}
\xi=CD_{1,1}(x-CRPIX1) + CD_{1,2}(y-CRPIX2)
\end{equation}
\begin{equation}
\label{e2}
\eta=CD_{2,1}(x-CRPIX1)+CD_{2,2}(y-CRPIX2).
\end{equation}
where the CRPIX1 and CRPIX2 represent the pixel coordinates of the centre. Then, the proper motions in units of mas yr$^{-1}$ ($d\xi$ and $d\eta$) are obtained from the first derivatives of Eqs. \ref{e1} and \ref{e2} where $dx$ and $dy$ represent the proper motions in units of pix yr$^{-1}$ as follows:
\begin {equation}
d\xi=CD_{1,1}dx + CD_{1,2}dy
\end{equation}
\begin{equation}
d\eta=CD_{2,1}dx+CD_{2,2}dy.
\end{equation}
In this study $d\xi$ and $d\eta$ correspond to $\mu_\alpha cos(\delta)$ and $\mu_\delta$, respectively, because the proper motion results are $\sigma$-clipped averages effectively at the tile centre, which has the right ascension ($\alpha$) aligned along $\xi$ and the declination ($\delta$) along $\eta$.  

The overall proper motion for SMC and 47 Tuc stars was calculated as described below. The same procedure was applied also to the background galaxies in order to correct for residual systematic effects and the results are discussed in Sect.~\ref{systematics}. Note that observations of sources in the overlapping regions of detectors are treated separately and these will result in multiple entries in the proper motion catalogues. Most of the sources appear on two, but different, detectors. Since they represent independent proper motion measurements, they contribute to reducing the statistical uncertainties of the resulting medians.

\begin{landscape}
\begin{table*}
\caption{Proper motions of different groups of sources in the tile SMC $5\_2$}
\tiny
\label{tilepm}
\[
\begin{array}{lrrrrrrrr|rrrrrrrr|rrrrrrrr}
\hline \hline
\noalign{\smallskip}
\multicolumn{9}{c|}{\mathrm{SMC}} & \multicolumn{8}{c|}{\mathrm{Mixed}} & \multicolumn{8}{c}{\mathrm{47\,Tuc}} \\
\mathrm{CMD}  & N^1 & \multicolumn{3}{c}{\mu_\alpha\cos(\delta)}  & N^1 & \multicolumn{3}{c|}{\mu_\delta}  & N^1 & \multicolumn{3}{c}{\mu_\alpha\cos(\delta)}  & N^1 & \multicolumn{3}{c|}{\mu_\delta}  & N^1 & \multicolumn{3}{c}{\mu_\alpha\cos(\delta)}  & N^1 & \multicolumn{3}{c}{\mu_\delta} \\
 & & \multicolumn{3}{c}{\mathrm{(mas\,yr^{-1})}} & & \multicolumn{3}{c|}{\mathrm{(mas\,yr^{-1})}} & & \multicolumn{3}{c}{\mathrm{(mas\,yr^{-1})}} & & \multicolumn{3}{c|}{\mathrm{(mas\,yr^{-1})}} & & \multicolumn{3}{c}{\mathrm{(mas\,yr^{-1})}} & & \multicolumn{3}{c}{\mathrm{(mas\,yr^{-1})}} \\   
 & & \mu & \sigma_\mu & \sigma & & \mu & \sigma_\mu & \sigma & & \mu & \sigma_\mu & \sigma & & \mu & \sigma_\mu & \sigma & & \mu & \sigma_\mu & \sigma & & \mu & \sigma_\mu & \sigma \\   
\hline
\noalign{\smallskip}
\mathrm{A} &     19  & +2.04  &     5.14  &     17.87      &    19  &    -5.17   &    4.11    &   14.31 & - & -& -&- &- &- &- &- &- & -& -& -& -& -& - & -\\
\mathrm{B} &     16  & +1.08  &   10.53      &  33.63     &     15   &  13.53    &   13.10   &    40.49 &- &- &- &- &- &- &- & -& -& -& -&- &- &- &- & -\\
\mathrm{C} &   205 & -0.39   &    1.48    &    16.87     &    210  &   -2.85   &    1.42    &   16.43 &- &- &- & -& -& -& -& -&- &- &- &- &- &- &- & -\\
\mathrm{D} &   898 & +3.90  &    0.72 &      19.26  &      912 &    -1.23   &    0.83   &    19.93 &- & -& -& -& -& -& -& -& -& -& -& -& -& -& -& -\\
\mathrm{E} & 5787 & +1.15  &     0.16   &    9.84    &    5879  &   -0.80   &   0.17   &    10.21 &  - & -& -& -& -& -& -& -& -& -& -& -& -& -& -& -\\
\mathrm{E}_1& - & -& -& -&- &- & -& -& 3048 &    +4.59   &   0.39   &   17.26 & 3100  & -1.50 & 0.43  &  19.13 & 325  &  +3.17  & 1.32  & 18.96 & 319 &   +0.04 & 1.35 & 19.19 \\
\mathrm{E}_2 &- & -& -& -&- &- & -& -& 2271 &    +2.14   &   0.17  &    6.32  & 2287  &    -0.54   &   0.18  &  7.05 &- & -& -& -&- &- & -& -\\ 
\mathrm{F} & 3615 & +9.89  &    0.23   &  11.16   & 3448  &    -1.04   &   0.21 & 10.07 & 3995 & +10.44  & 0.19  &  9.66  & 3870  &  -1.21  &  0.16  & 8.20 & 562 & +10.47  & 0.41  & 7.82   & 571  &  -1.46  &  0.40  & 7.68\\
\mathrm{G} &       2 & +11.31 &    2.59    &    2.93     &      2   &   +6.81   &    8.50   &    9.60 &- & -& -&- &- & -& -&- &- &- &- & -&- &- &-& - \\
\mathrm{H} &  698 & +7.31 &    0.23    &   4.88    &     708  &    -1.12  &    0.22   &    4.77 &- & -& -&- &- &- &- &- &- &- &- & -&- &- & -& -\\
\mathrm{I} & 1536 & +4.32  &    0.17    &   5.28   &     1611    &  -1.61  &    0.17  &   5.44 & -& -& -& -&- & -& -& -&- & -&- &- &- &- &- & -\\
\mathrm{J} & 6003 & +2.20  &    0.13      &  8.12    &    6004  &   -0.91  &    0.14     &  8.49 & 8375   &    +3.64   &   0.12  &   8.98   &     8201  &    -0.63  &    0.12  &   8.77 & 9088   &    +6.79  &     0.11    &   8.39    &   9295  &    -0.84   &   0.12    &   9.15\\
\mathrm{K} & 2684 & +0.95 &   0.08     &   3.43   &     2843  &   -0.51  &   0.09 &      3.64 & 2958    &     +2.10  &    0.08   &    3.59   &     3381 &    -0.78  &    0.09  &      4.15 & 871 &  +2.16  &    0.17   &    3.97     &   947  &    -0.12   &    0.17  &      4.15\\
\mathrm{L} & 316 &  +6.27  &     0.93 &      13.24    &     313  &    -0.03  &    0.98   &    13.90 & 599  &     +8.61    &   0.83    &   16.22  &       526  &   -1.01  &     0.61   &    11.11 & 263 &   +7.25  &     0.94   &    12.11  &       256    &  -4.44   &    0.93 &      11.93\\
\mathrm{M}&- & -&- &- & -& -&- & -& 8819   &    +7.08  &    0.07  &     5.04   &     8910   &   -1.70   &   0.07     &  5.26 & 23404  &     +7.32 &    0.04  &   5.16 &      24050  & -1.15 &    0.05 &      5.77 \\
\mathrm{N}& -& -&- &- & -&- & -& - & 5515   &    +4.52  &    0.18  &     10.52   &     5455   &   -0.64  &    0.18   &    10.81 & 11979   &    +7.25  &    0.12  &     10.55  &     12000  &   -0.84   &   0.12    &    10.91\\
\mathrm{O}&- &- &- &- & -&- &- &  -& 8895  &    +9.35  &    0.23  &     17.04   &    8881  &    -1.71    &  0.24    &    18.42 & 13911   &    +8.05  &    0.15  &      13.89    &   13979   &   -0.67  &    0.16  &     14.70\\
\noalign{\smallskip}
\hline
\hline
\end{array}
\]
$^1$ About $50\%$ are unique sources (see Sect.~\ref{pmsec} for details).
\end{table*}
\end{landscape}

\subsection{SMC area}
The CMD of sources contained in the SMC area ($\rho>40^\prime$) was divided into $12$ regions identified with letters ABCDEFGHIJKL and enclosing populations of different types and ages (Fig.~\ref{pmcmd} left). These CMD regions were defined as in \citet{2014A&A...562A..32C} for the LMC and were adjusted to the different mean distance, extinction, and metalicity of the SMC by applying a shift in magnitude, $K_\mathrm{s}=+0.6$, and in colour, $(J-K_\mathrm{s})=-0.04$ \citep{2000A&A...359..601C}. The shift in magnitude is slightly larger than it appears in the latest studies, e.g. from \citet{2014AJ....147..122D} and \citet{2015AJ....149..179D} we obtain $\sim0.5$ mag, but it does not influence our investigation. The proper motion of each region was derived from the median of the proper motion of the stars enclosed in the region after the application of a sigma clipping procedure. Sources with proper motion values further than $3\sigma$ from the median were rejected and a new median value was calculated from the remaining sources; this procedure was repeated a maximum of ten times or until only a handful of sources were removed in a given iteration. The distinction between SMC, 47 Tuc, and MW stars is reflected in their proper motions. Table \ref{tilepm} shows the proper motion derived for each CMD region. In particular, Col. $1$ lists the CMD region, Cols. $2-5$ list the number of measurements, which as explained above is $\sim50\%$ larger than the number of sources, their median proper motion, uncertainty and dispersion, respectively, for the $\mu_\alpha\cos(\delta)$ component, and Cols. $6-9$ list the same information for the $\mu_\delta$ component.

\subsection{47 Tuc area}
The CMD of sources comprised within the 47 Tuc area ($\rho<20^\prime$) was also divided into separate regions, but these differ in part from those of the SMC region because of the distribution of the 47 Tuc sources. The exact boundaries of each region are given in Table \ref{tucregions}. The middle panel of Fig.~\ref{pmcmd} shows the distribution of sources in the 47 Tuc area. Regions J and K are as in the SMC area while regions E$_1$ and N are the faint and bright part of region E of Fig.~\ref{pmcmd} (left). Regions M and O are newly defined and consequently, regions F and L are also modified. Regions E$_1$ and K contain mostly SMC stars while regions F and L contain mostly MW stars. Regions M and O contain mostly 47 Tuc stars (red clump, RGB, SGB, turn-off, and main sequence stars). Regions J and N contain a large number of 47 Tuc main sequence stars and many SMC stars. The proper motion of stars enclosed in each region was calculated following the same procedure as for the stars in the SMC area. Proper motions for each region are listed in Table \ref{tilepm}. In particular, Cols. $10-13$ list the number of measurements where $\sim50\%$ are unique sources, their median proper motion, uncertainty and dispersion, respectively, for the $\mu_\alpha\cos(\delta)$ component and Cols. $14-17$ list the same information for the $\mu_\delta$ component.

\subsection{Mixed area}
\begin{figure*}
\resizebox{\hsize}{!}{\includegraphics{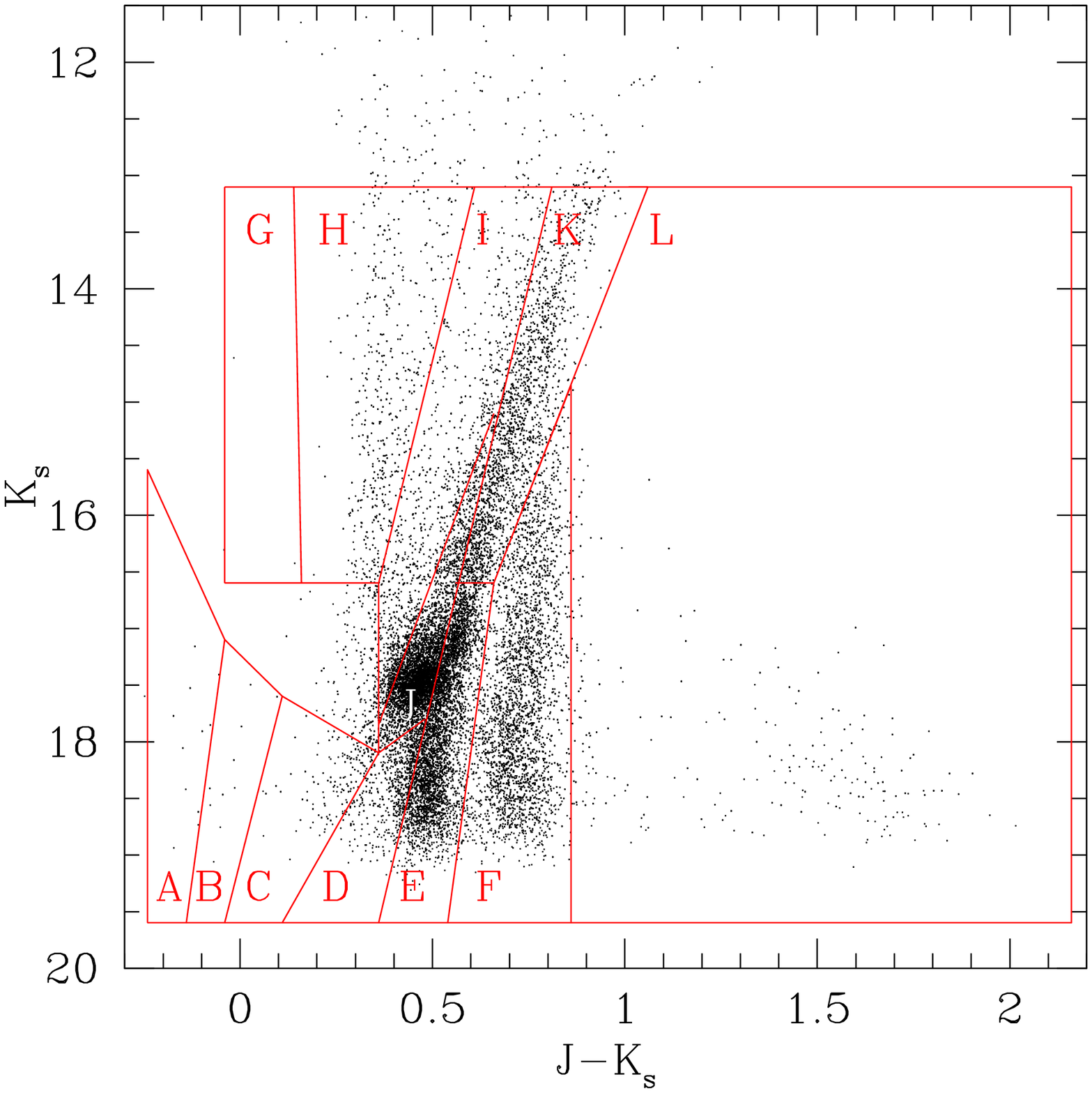}
\includegraphics{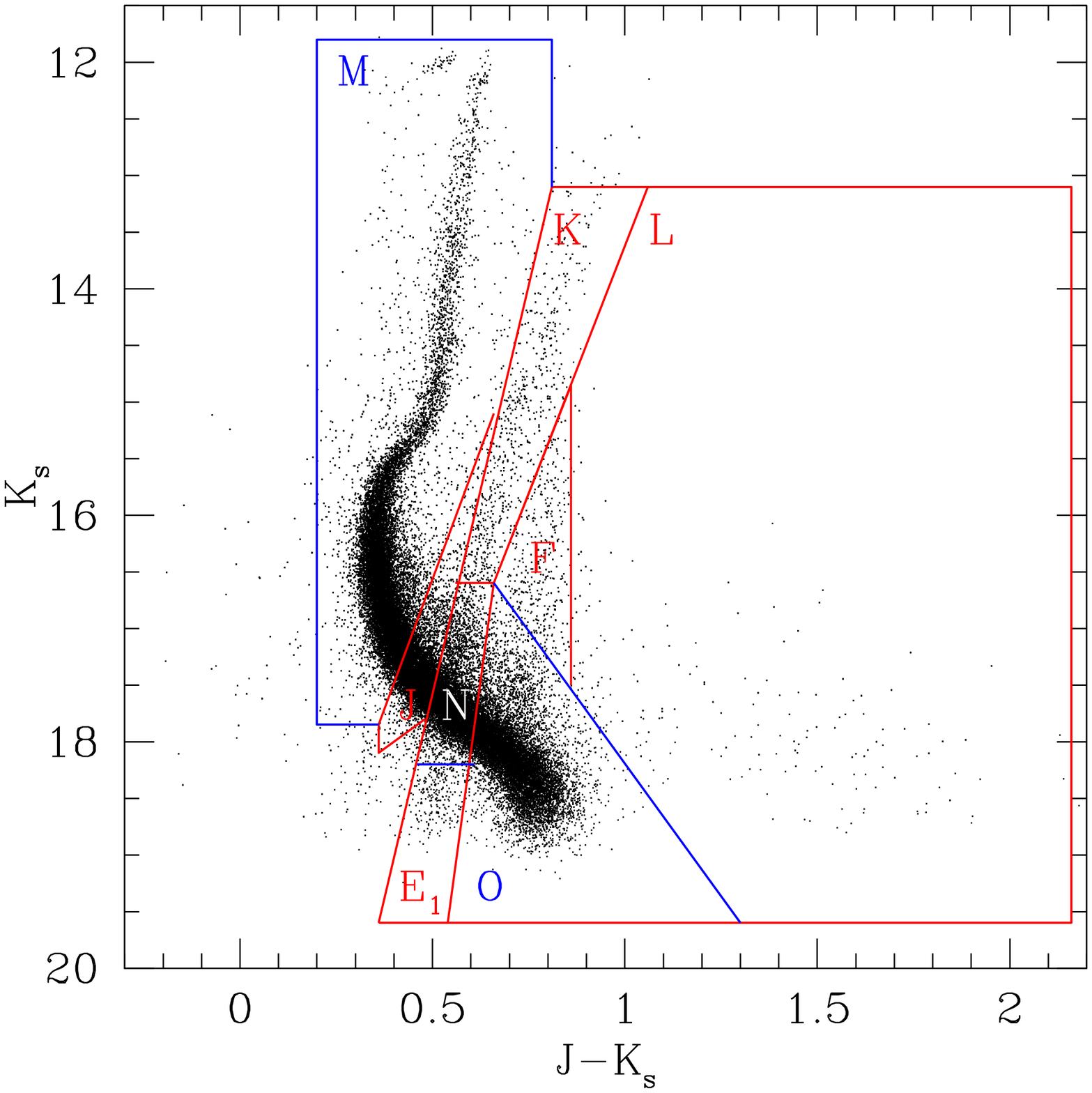}
\includegraphics{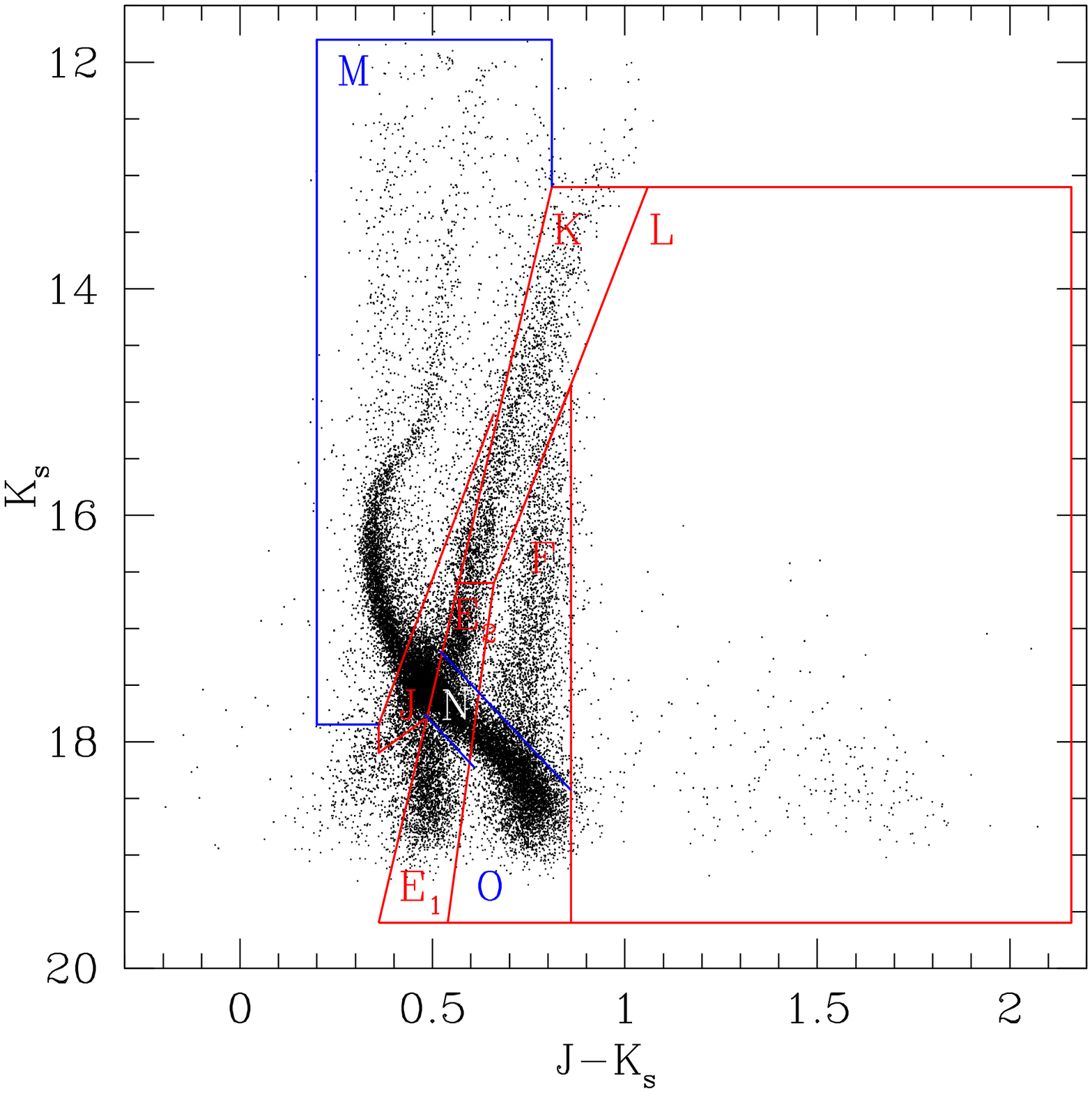}}
\caption{CMDs of star-like sources, with measured proper motions, in tile SMC $5\_2$ that are located a distance $\rho>40^\prime$ (left), $20^\prime<\rho<40^\prime$ (middle), and $20^\prime<\rho<40^\prime$  (right) from the 47 Tuc centre. Different CMD regions enclose sources with a different mean age. In the left panel, regions are defined as in \citet{2014A&A...562A..32C}, shifted both in magnitude and colour to account for the distance, metallicity and reddening of the SMC. Most of the sources belong to the SMC except for sources in regions F, H, and L which most likely belong to the MW and/or 47 Tuc. 
In the middle panel, regions are in part different to account for the contribution of the 47 Tuc stars. Most of the sources belong to the cluster except for sources in region K, that most likely belong to the SMC, and for sources in regions F and L, that may belong to the MW. In the right panel, the prevalence of SMC, 47 Tuc, and MW sources within different regions can be judged in comparison with the other panels.}
\label{pmcmd}
\end{figure*}

The CMD of sources comprised in the mixed area ($20^\prime<\rho<40^\prime$) was divided into separate regions that mostly resemble those defined in the 47 Tuc area (compare Fig.~\ref{pmcmd} middle and right panels). In particular, regions JKM were unaltered, but region N was reduced and as a consequence region E$_1$ was extended to brighter magnitudes, and region E$_2$ was introduced to better distinguish SMC stars brighter than 47 Tuc main-sequence stars. The exact boundaries of each region are given in Table \ref{mixedregions}. Regions F and O were modified to better distinguish MW stars brighter than the 47 Tuc main-sequence stars, while region L was restored as in the SMC area (Fig.~\ref{pmcmd} left). The right panel of Fig.~\ref{pmcmd} shows the distribution of sources in the mixed area. The proper motions of stars enclosed in each region show good agreement with those derived previously in analogous regions of the CMD. Proper motions for each region are listed in Table \ref{tilepm}. In particular, Cols. $18-21$ list the number of measurements where $\sim50\%$ are unique sources, their median proper motion, uncertainty and dispersion, respectively, for the $\mu_\alpha\cos(\delta)$ component and Cols. $22-25$ list the same information for the $\mu_\delta$ component.

 \subsection{Errors}
 \label{systematics}
To evaluate the accuracy of the proper motion measurements it is important to establish sources of possible random and systematic uncertainties. The proper motion of each galaxy was derived following the same procedure applied to stars (Sect.~\ref{pmsec}) and their resulting median proper motion was $(\mu_\alpha\cos(\delta), \mu_\delta)=(-0.45\pm0.12, -0.15\pm0.12)$ mas yr$^{-1}$ with a dispersion of about $14.4$ mas yr$^{-1}$. This value corresponds to $\sim22500$ galaxies distributed across the tile. It is sufficiently close to zero to indicate that the reference system they define is not moving, but it is also evidence of a residual systematic uncertainty. The proper motions of SMC, 47 Tuc, and MW stars, discussed in this paper, were corrected for it by subtracting from their respective proper motions that of the galaxies.

Ground-based astrometric measurements are limited by atmospheric turbulence \citep{1995PASP..107..399H}. The typical {\it rms} size of this random effect depends on the seeing, the separation between objects ($\theta$), and the exposure time ($T$) as follows \citep{2013A&A...554A.101B}:
\begin{equation}
\sigma = 38 (\mathrm{mas})\times(\frac{\theta}{30^\prime})^\frac{1}{3}\times \frac{1}{\sqrt{T(\mathrm{sec})}}.
\end{equation}
We have calculated the average separation between galaxies and stars in the detector images of the mixed region ($20^\prime<\rho<40^\prime$ from the 47 Tuc centre). The latter samples well the different stellar populations in the tile across multiple detectors. We obtain $\sim 1.5^\prime$ separation that combined with a detector exposure time of $5$ (s) $\times15$ (exposures) $\times5$ (jitter positions)$ =375$ s \citep{2011A&A...527A.116C} results in an rms of $\sim 0.55$ mas. 

Atmospheric differential refraction influences the measurement of the centroid of objects \citep{1982PASP...94..715F}. In this work we use observations obtained within a narrow range of air masses and with the same $K_\mathrm{s}$ filter to limit these positional errors. The relative position of objects with a different mean colour, however, may present differential atmospheric refraction effects. These are difficult to characterise because they require sampling of the same type of objects at different colours, but it is not often the case that objects of a given type are distributed in the CMD within a sufficiently large range of colours to cover the full extent of the measurements that are being compared, including reference objects. We have examined the colour distribution of the proper motions of galaxies, 47 Tuc stars (region M of Fig.~\ref{pmcmd} middle panel), and SMC stars (region K of Fig.~\ref{pmcmd} left panel) which together sample a range of about 2 mag in colour. We found no significant trend within either of their distributions indicating that possible atmospheric differential refraction effects are small (Figs.~\ref{colfigs} and \ref{colfigsbin}). Furthermore, any residual non-linearity of the difference between the differential atmospheric refraction at the various epochs is also minimised and can also be well approximated using a linear differential mapping.  The absence of any measurable colour dependence for the derived proper motions for the different objects studied demonstrates the veracity of this approach (Sect.~\ref{pmsec}).

\begin{figure}
\resizebox{\hsize}{!}{\includegraphics{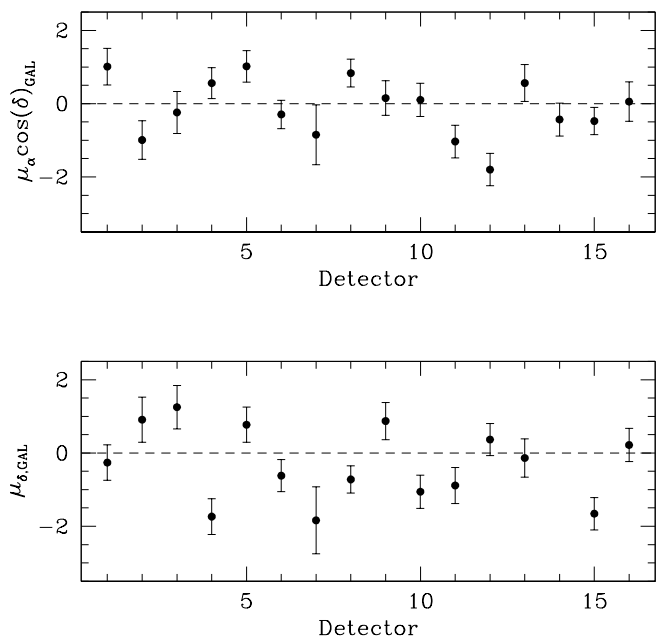}}
\caption{Proper motions of galaxies in mas yr$^{-1}$ as a function of detector number.}
\label{galdet}
\end{figure}

We recall that individual VISTA detector images are calibrated independently with 2MASS stars, so for each epoch we have $16$ (detectors) $\times6$ (pawprints) $=96$ independent measurements, and the systematic errors are dominated by the 2MASS uncertainties. Figure \ref{galdet} shows the distribution of the proper motions derived for the galaxies as a function of detector number, after correcting for a residual systematic effect described above. We found an average scatter of $\sim0.7$ mas yr$^{-1}$, one outlier (detector \#14 in the $\mu_\delta$ component), but not a trend. The large error bar associated with detector \#$7$ is due to the presence of the 47 Tuc cluster and a reduced number of galaxies with respect to the other detectors.  The effect of outliers would not influence the well-sampled population distributed across multiple detectors. With respect to Fig.~\ref{map} detector \#1 is in the top-right corner and detector \#16 in the bottom-left. Detector numbers increase from top to bottom and from right to left. Note that because of the large gaps among the VISTA detectors objects are also observed in at least one of the adjacent detectors. In particular, objects in the three areas shown in Fig.~\ref{map} are observed in all detectors except for the following: \#3, \#6, \#7, \#8, and \#11 for the SMC area (i.e. excluding a cross-like region centred on 47 Tuc), \#7 and \#13 for the mixed area (i.e. excluding the centre of 47 Tuc and the top-left corner of the tile), and \#1, \#5, \#9, \#13, \#14, \#15, and \#16 for the 47 Tuc area (i.e. excluding the top and left borders of the tile). Taking these detector numbers into account we do not find any systematic effect that may suggest a bias in the reference frame calibration of the different populations across the tile (Fig.\ref{map}).

Furthermore, we examined the proper motions of background galaxies as a function of right ascension ($\alpha$) and declination ($\delta$). Results are shown in App.~\ref{pmtrends}. In both directions and for both components the proper motions are consistent with zero, except for galaxies located in the northermost region of the tile.  This area is sampled by detectors \#1, \#5, \#9, and \#13. Detector \#1  contains a series of bad pixels grouped in three large areas and it is possible that the VMC five-sequence jitter pattern is not sufficient to remove their effect on many of the extracted sources. Detector \#4 has a number of bad rows that are not always corrected very well by the standard image processing. For more details about known issues associated with the VISTA detectors see on the Cambridge Astronomy Survey Unit webpages\footnote{http://casu.ast.cam.ac.uk/surveys-projects/vista/technical/known-issues}. In our {\it ppErrBits} flags selection we have chosen to keep sources that have at least one bad pixel in the default aperture (Sect.~\ref{data}), these are in most cases good quality detections. We note that most of the stars in 47 Tuc and in the SMC are located in the bottom $2/3$ of the tile where the best agreement among the proper motions of galaxies is achieved. The scatters of the proper motion measurements are comparable and correspond to $\sim0.65$ mas yr$^{-1}$.
We conclude that the uncertainty associated with the galaxies' proper motions, scaled with respect to the root of the number of measurements ($16$ detectors and $12-13$ points for the $\alpha$-$\delta$ trends) and averaged, $0.18$ mas yr$^{-1}$,  is a good estimate of the systematic errors in our analysis.

\section{Proper Motion Results}
\label{results}

\begin{figure*}
\resizebox{17cm}{11cm}{\includegraphics{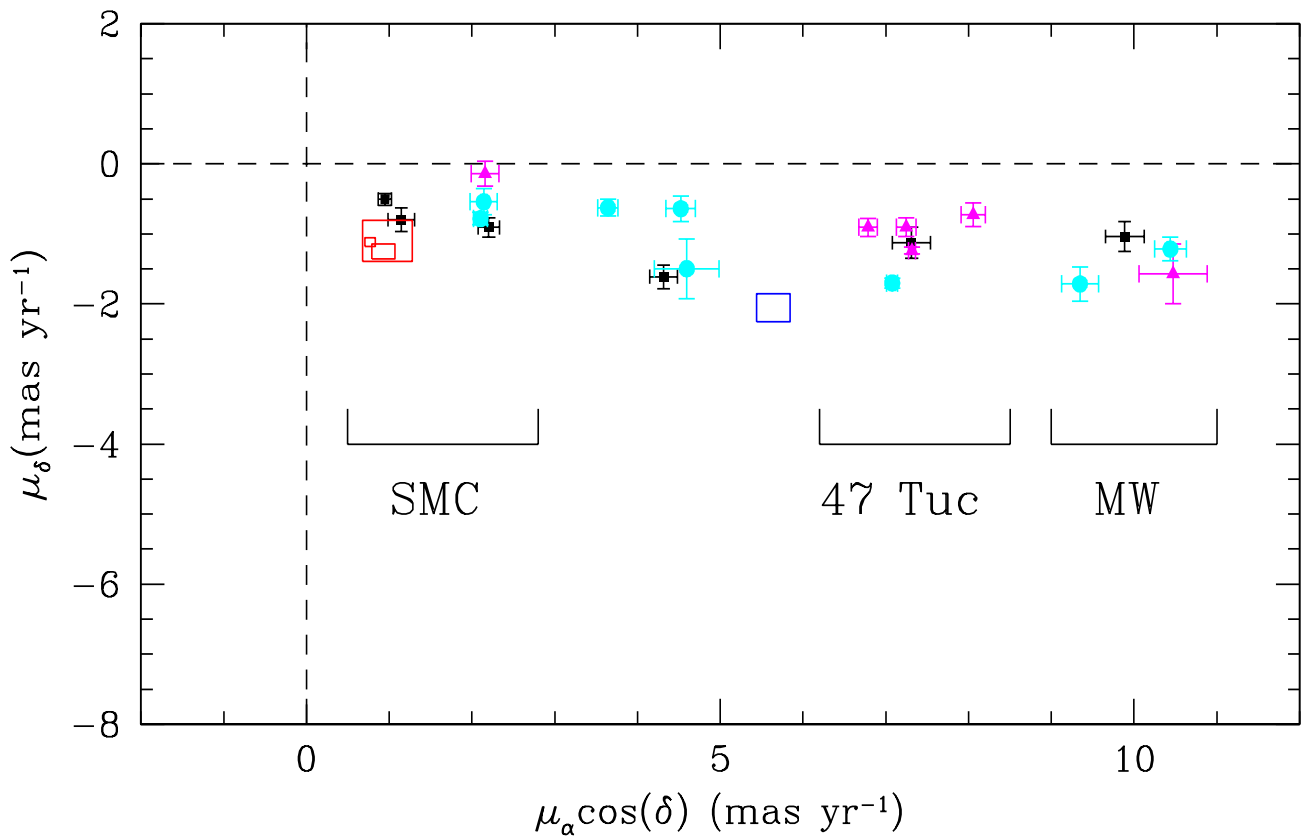}}
\caption{Proper motions of stars enclosed within different CMD regions (see Table \ref{tilepm} for details). Each point corresponds to the median of at least $500$ stars. Different colours indicate regions taken from areas dominated by SMC stars (black squares), 47 Tuc stars (magenta triangles), and mixed stars (cyan dots). Centre of mass proper motion measurements from the literature for both the SMC and 47 Tuc are indicated by empty squares whose sizes correspond to their statistical uncertainties. For the SMC (red) they are from \citet{2013ApJ...764..161K, 2011AJ....141..136C, 2010AJ....140.1934V} in increasing order of rectangular size and for 47 Tuc (blue) it is from \citet{2003AJ....126..772A}.
}
\label{pmfig}
\end{figure*}

The proper motions of sources located in the different regions of the CMD are shown in Fig. \ref{pmfig}; for clarity, only regions with proper motion uncertainties $<0.5$ mas yr$^{-1}$ are indicated.

\subsection{SMC area}

The points associated with the area dominated by SMC stars show that stars populating the red clump (J) have a proper motion that differs from that of RGB stars above (K) and below (E) it, that are similar to each other. This difference may be due to population effects, in \citep{2014A&A...562A..32C} the proper motion was shown to decrease with increasing age of the stellar population, or to the presence of MW and/or 47 Tuc stars within region J. The latter is not excluded because MW and 47 Tuc stars are present in regions D and H that are located at faint and bright magnitudes of region J, respectively. 
We have checked the contamination of region J by calculating the proper motion for two subsets of stars distributed at high ($\delta>-71.93^\circ$) and low ($\delta<-71.93^\circ$) declinations. We found that those towards the SMC, about $3200$ stars, show a median proper motion of $(\mu_\alpha\cos(\delta), \mu_\delta)=(+1.11\pm0.17, -1.58\pm0.18)$ mas yr$^{-1}$ and those away from it, about $2800$ stars, show $(\mu_\alpha\cos(\delta), \mu_\delta)=(+2.89\pm0.20, +0.07\pm0.22)$ mas yr$^{-1}$. The latter is clearly influenced by MW and/or 47 Tuc stars while the former agrees with the results obtained for RGB stars. 
For the giants at bluer colours and adjacent to the RGB (DI) the influence of some 47 Tuc and MW stars shifts their average proper motion to larger value resulting eventually in a larger uncertainty. There are not enough stars in the regions tracing the young population of the SMC (A, B, C, G) for a meaningful comparison.  On the other hand, stars in region H, which is mostly composed of 47 Tuc stars (Fig. \ref{pmcmd}), show a proper motion clearly different from that of SMC stars. MW stars (F) are noticeably different from both SMC and 47 Tuc stars at proper motions of $\sim10$ mas yr$^{-1}$ in $\mu_\alpha\cos(\delta)$. Region L, which is occupied mostly by background galaxies, contains MW stars of late spectral type; their proper motion is far from zero. Only proper motions corresponding to regions EFHIJK are shown in Fig.~\ref{pmfig}.

\subsection{47 Tuc area}

The points associated with the area dominated by 47 Tuc stars indicate an uncontaminated RGB and upper main sequence (M) proper motion around $7.32$ mas yr$^{-1}$ in $\mu_\alpha\cos(\delta)$ and $-1.15$ mas yr$^{-1}$ in $\mu_\delta$ with very small uncertainties ($0.05$ mas yr$^{-1}$) which is also in excellent agreement with the proper motion obtained from the lower main sequence stars (N) despite the presence of SMC stars in the region. On the other hand, the proper motion of lower main sequence stars (O) is slightly higher and that can be attributed to a contamination by MW stars. We also found no significant difference between the proper motion of stars of region M that are brighter and fainter than the main sequence turnoff, respectively.  For region K and E$_1$, dominated by SMC stars, we find a higher proper motion compared to that derived above, this is further discussed in the next section. The region where the largest density of SMC stars is found (J) shows only a minor effect on the considerably more numerous population of 47 Tuc main sequence stars. The proper motion of MW stars (FL) reflects the value for stars in a similar CMD region but further from the cluster centre (see above) although stars in the faintest part are in this case excluded. The somewhat larger proper motion may be due to having more thin than thick disk stars in populating the region or vice versa. Only proper motions corresponding to regions FJKMNO are shown in Fig.~\ref{pmfig}.

\subsection{Mixed area}

The region where a mixed population of 47 Tuc and SMC stars is expected shows proper motions where the contribution of one or the other population are clearly distinct. The proper motion of the MW population (F) is in excellent agreement with the previous determinations. The brightest 47 Tuc populations (M) do not seem to be influenced by a contamination from either SMC or MW stars. On the contrary, where SMC stars are numerous (N) the proper motion shifts to smaller $\mu_\alpha\cos(\delta)$ values and where the MW stars are numerous (O) the proper motion shifts to larger $\mu_\alpha\cos(\delta)$ values. The predominance of SMC red clump stars over the 47 Tuc main sequence (J) and of the latter over the SMC RGB (E$_1$) results in proper motions that are somewhat shifted from those found for the SMC area. The faint RGB below the 47 Tuc main sequence (E$_2$) supports SMC-like proper motions. The proper motion of region K is somewhat higher than that derived in the same region of the SMC area. We have checked the contamination of MW stars in region K (see also App.~\ref{pmtrends}) by calculating the proper motion for the subset of stars that are distributed on each side of the 47 Tuc cluster. We found that those towards the SMC, about $2300$ stars, show a median proper motion of $(\mu_\alpha\cos(\delta), \mu_\delta)=(+1.37\pm0.09, -1.08\pm0.10)$ mas yr$^{-1}$ and those towards the opposite direction, about $900$ stars, show $(\mu_\alpha\cos(\delta), \mu_\delta)=(+3.00\pm0.20, -0.46\pm0.18)$ mas yr$^{-1}$. The latter is clearly influenced by MW stars while the former is in line with the results obtained for SMC stars. Proper motions in all regions, except L, are shown in Fig.~\ref{pmfig}.

\begin{table}
\caption{Median proper motions in the tile SMC $5\_2$.}
\tiny
\label{avgpm}
\[
\begin{array}{@{}lrrrrrrrr@{}}
\hline \hline
\noalign{\smallskip}
\mathrm{CMD}  & N^1 & \multicolumn{3}{c}{\mu_\alpha\cos(\delta)}  & N^1 & \multicolumn{3}{c}{\mu_\delta} \\
\mathrm{Region} & & \multicolumn{3}{c}{\mathrm{(mas\,yr^{-1})}} & & \multicolumn{3}{c}{\mathrm{(mas\,yr^{-1})}} \\   
 & & \mu & \sigma_\mu & \sigma & & \mu & \sigma_\mu & \sigma \\   
\hline
 \noalign{\smallskip}
 \mathrm{SMC_{Red}} & 10328   &    +1.16 &      0.07 &      5.93 &      10626 &    -0.81 &      0.07 &       6.17\\
 \mathrm{47\,Tuc} & 69862  &     +7.26  &   0.03  &     6.90  &     71195 &    -1.25  &   0.03  &     7.49 \\
 \mathrm{MW} &  8097   &       +10.22 &      0.14 &       9.95 &       7770  &    -1.27 &     0.12 &       8.70\\
 \mathrm{NGC\,121} & 263 & +2.75 & 0.26 & 3.31 & 374 & -4.67 & 0.47 & 7.20 \\
% \mathrm{Galaxies}^2 & 22466 & -0.45 &  0.12  &      14.10  &     22385 &     -0.15 &     0.12 &      14.61 \\
\noalign{\smallskip}
\hline
\hline
\end{array}
\]
$^1$ About $50\%$ are unique sources (see Sect.~\ref{pmsec} for details).

%$^2$ The proper motion of galaxies has been subtracted from the stellar values listed above (see Sect.~\ref{pmsec} for details).}
\end{table}

\subsection{Combined CMD regions}

The proper motions of SMC stars, in CMD regions with a small, if any, contribution by MW and 47 Tuc stars, are consistent with each other (red clump and RGB stars). The proper motions of similar stars at larger distances from the SMC appear influenced by MW/47 Tuc stars unless projection effects within the tile are larger than expected. The proper motion range spanned by 47 Tuc stars is $\sim1$ mas yr$^{-1}$ and in this case it may  be related to stellar components distributed differently as a function of cluster radius. A detailed analysis of the internal kinematics of 47 Tuc stars, using these proper motion data, is presented in a forthcoming paper.

Table \ref{avgpm} lists the median proper motions for the SMC, 47 Tuc, and MW components derived from the combination of all of the independent proper motion measurements in the following regions.  For 47 Tuc we combined regions JMNO from Fig.~\ref{pmcmd} (middle), regions MO from Fig.~\ref{pmcmd} (right), and region H from Fig.~\ref{pmcmd} (left); we used in total about $70000$ measurements. For the MW we combined regions F from all three panels of Fig.~\ref{pmcmd} using in total about $8000$ measurements. For the SMC we combined regions EK and J ($\delta<-71.93^\circ$ only, i.e. closer to the SMC) from Fig.~\ref{pmcmd} (left) totalling about $10500$ measurements and resulting in a proper motion representative of its red stellar population (red clump and RGB stars).

The proper motion of NGC 121 (Sect.~\ref{smccluster}) is also listed in Table \ref{avgpm}. In particular, Col. $1$ lists the CMD region, Cols. $2-5$ list the number of measurements where $\sim50\%$ are unique sources, the median proper motion, its uncertainty and dispersion for the $\mu_\alpha\cos(\delta)$ component, respectively, and Cols. $6-9$ list the same information for the $\mu_\delta$ component.

\subsection{Comparison to previous measurements}
In this section we compare our proper motion results with those from recent studies. We note that the comparison refers always to proper motion measurements that are not corrected for solar reflex motion. The most accurate proper motion of the SMC to date has been measured using the HST \citep{2013ApJ...764..161K}. The authors analysed five fields centred on one background quasar and aligned with the East-West axis of the galaxy at an angle of $\sim45^\circ$. Their observations referred to a time span of $3-7$ yr, but comprised a relatively small number of objects. By assuming a simple idealistic model for the geometry of the galaxy to correct for viewing perspective effects, they obtained a centre of mass proper motion of  $(\mu_\alpha\cos(\delta)$, $\mu_\delta)=(+0.772\pm0.063$,$-1.117\pm0.061)$ mas yr$^{-1}$ (Fig.~\ref{pmfig}). Perspective effects are $\sim0.1$ mas yr$^{-1}$, but can be larger at larger distances from the centre. Compared to our measurements at $\sim3^\circ$ from the HI centre \citep{2004ApJ...604..176S}, from different SMC populations, we find a difference of $0.4$ mas yr$^{-1}$ in both components. This difference may be attributed to the combination of viewing perspective, which is not corrected for in our study, and of systematic uncertainties, but it may also suggest a difference between the proper motion of the stellar populations observed with HST and VISTA. Systematic uncertainties are discussed in Sect.~\ref{systematics}, but there may be an additional effect due to the calibration using quasars and galaxies because the two types of objects have different mean colours \citep[see e.g.][]{2013A&A...549A..29C}. 
Our proper motion for the SMC$_\mathrm{Red}$ is dominated by stars in the red clump and RGB phases which typically have ages between $1$ Gyr and a few Gyr. On the other hand, HST observations are dominated by main sequence stars with a small number of SGB, red clump and RGB stars \citep{2006ApJ...638..772K, 2008AJ....135.1024P}. 

The proper motion of the SMC was derived by \citet{2011AJ....141..136C} from observations with the Ir\'{e}ne\'{e} du Pont 2.5 m telescope at Las Campanas Observatory (Chile) covering five fields of $9^\prime \times 9^\prime$ each, also centred on a background quasar. They measured $(\mu_\alpha\cos(\delta)$, $\mu_\delta)=(+0.93\pm0.14$,$-1.25\pm0.11)$ mas yr$^{-1}$ (Fig.~\ref{pmfig}) and did not detect population-dependent internal motions within two of their fields, those with a sufficient number of blue and red stars. They suggested, however, that fields to the outer East and South of the SMC may be affected by streaming motions.  Their proper motion differs from that of the other fields, to the North and West of the galaxy, by $\sim 1$ mas yr$^{-1}$ in $\mu_\delta$.

Oxygen-rich AGB stars analysed by \citet{2010AJ....140.1934V}, from observations included in the Southern Proper Motion program \citep{1974IAUS...61..201W}, result in proper motion values of $(\mu_\alpha\cos(\delta)$, $\mu_\delta)=(+0.98\pm0.30$,$-1.10\pm0.29)$ mas yr$^{-1}$ (Fig.~\ref{pmfig}). Their sample comprises about $1000$ stars distributed more or less homogeneously over the galaxy; the spatial resolution was too coarse to sample the main central body of the SMC. Furthermore, by averaging the contributions of all stars, any potential asymmetric effect associated with the tidal interaction with the LMC would not be detected. In our study AGB stars are brighter than the regions analysed in Fig.~\ref{pmcmd}. Their number is not large and their analysis would result in large proper motion uncertainties. Moreover, the OGLE III and EROS-2 microlensing surveys that we could use to select long-period-variable stars (AGB stars), as was done in \citet{2014A&A...562A..32C}, marginally overlap with the VMC tile SMC 5\_2 and provide only a sample of $\sim100$ objects. This is also too small a number to obtain significant proper motion values and we postpone the analysis of the proper motion of the AGB component over the entire SMC to a subsequent study. 

The proper motion of stars in the core of the 47 Tuc cluster is shown in Fig.~\ref{pmfig} and refers to HST observations calibrated against the SMC background stellar population to the cluster; it corresponds to $(\mu_\alpha\cos(\delta)$, $\mu_\delta)=(+5.64\pm0.20$,$-2.05\pm0.20)$ mas yr$^{-1}$ \citep{2003AJ....126..772A}. The SMC value adopted to obtain the absolute proper motion of the cluster was $(\mu_\alpha\cos(\delta)$, $\mu_\delta)=(+0.92\pm0.20$,$-0.69\pm0.20)$ mas yr$^{-1}$ \citep{1999IAUS..192..409I}. In our study the difference between the 47 Tuc and SMC$_\mathrm{RED}$ proper motions, about $6$ and $0.4$ mas yr$^{-1}$  in $\mu_\alpha\cos(\delta)$ and $\mu_\delta$ respectively, is comparable to that derived by \citet{2003AJ....126..772A}.

\subsection{NGC 121}
\label{smccluster}
\begin{figure}
\resizebox{\hsize}{!}{\includegraphics{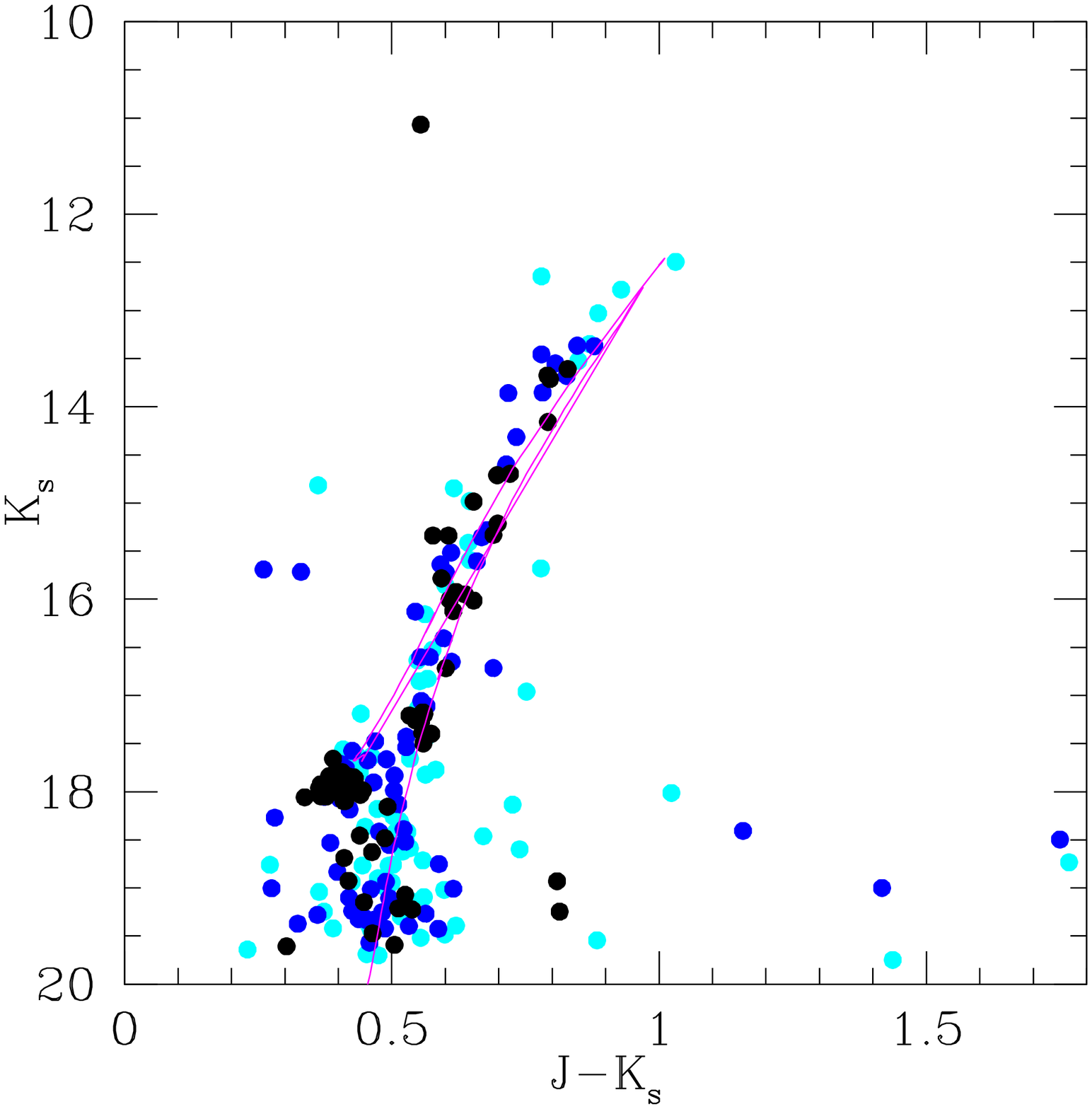}
\includegraphics{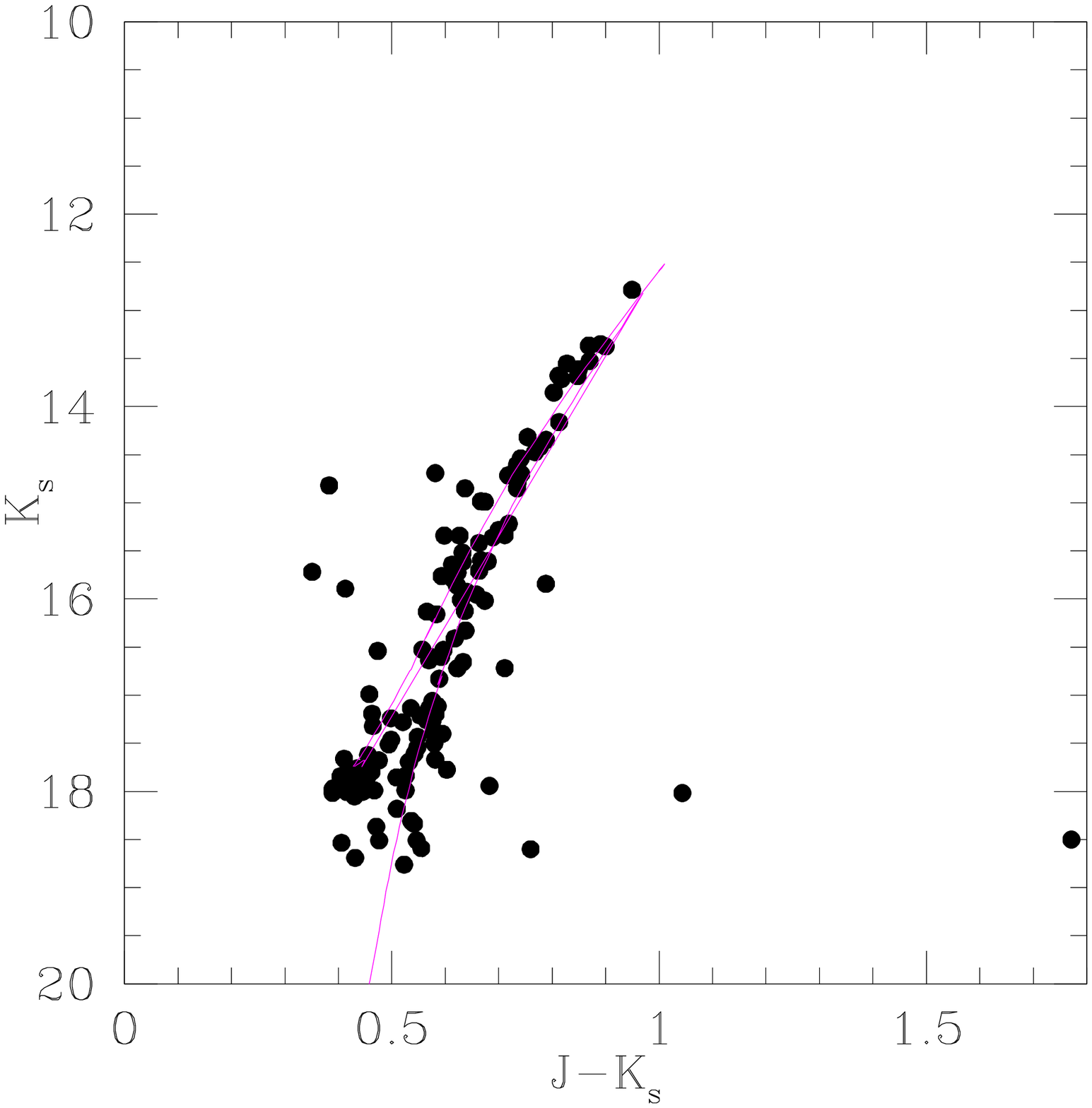}}
\caption{
The CMD of star-like sources within $1.9^\prime$ of the SMC star cluster NGC 121. In the left panel different colours indicate objects with a different probability of being cluster members: $100$\% (black), $75$\% (blue), and $50$\% (cyan). In the right panel only sources within $3\sigma$ from the cluster proper motion listed in Table \ref{avgpm}, and within the same radius, are shown. A PARSEC isochrone, assuming $(m-M)_0=18.9$ mag and $E(B-V)=0.04$ mag, indicates a $10.6$ Gyr old population with [Fe/H]$=-1.19$ dex.}
\label{ngc121}
\end{figure}

The SMC star cluster NGC 121 is located at approximately $35^\prime$ from the 47 Tuc centre.
Figure \ref{ngc121} (left) shows its CMD obtained from objects with a star-like shape located within $1.9^\prime$ of the cluster centre; it is characterised by a narrow RGB. Its size corresponds to the estimate by \citet{2008MNRAS.389..678B} and the cluster star membership of each object to the cluster has been estimated following the procedure described in \citet{2012MNRAS.425.3085P}. NGC 121 has an age of $10.6$ Gyr \citep{2001ApJ...562..303D} and a metallicity [Fe/H]$=-1.19\pm0.12$ dex \citep{1998AJ....115.1934D}. We superimposed in the figure a PARSEC \citep{2012MNRAS.427..127B} isochrone with these parameters and assuming a distance modulus $(m-M)_0=18.9$ mag and an extinction $E(B-V)=0.04$ mag. Note that it is beyond the scope of this work to revise the cluster age and metallicity, which will be performed in a separate study \citep[see e.g.][]{2014A&A...570A..74P}. 

In this study we derive the proper motion of NGC 121 by using all stars within $1.9^\prime$ of the cluster centre and from their median value we obtain $(\mu_\alpha \cos(\delta)$, $\mu_\delta)=(+2.75\pm0.26$,$-4.67\pm0.47)$ mas yr$^{-1}$. The large proper motion errors reflect the small number of sources, about $300$, included in the calculation (Table \ref{avgpm}) and the presence of intervening populations that cannot be disentangled with our procedure (see below). This value differs from that of the SMC and if this difference were confirmed, it could indicate that NGC 121 is not bound to the SMC. This highly speculative suggestion is, however, important in the context of the recently discovered satellite galaxies potentially associated with the Magellanic Clouds \citep[e.g.][]{2015ApJ...807...50B, 2015ApJ...805..130K, 2015ApJ...804L...5M}.

Figure \ref{ngc121} shows the CMD of NGC 121 including only sources within $3\sigma$ from the cluster proper motion (right). There are $231$ sources that satisfy this criterion. This number agrees with the result shown in left panel where we obtained that $239$ sources have a probability of being cluster members higher than $50\%$. The proper motion cleaned CMD shows a cluster RGB that is somewhat tighter than the one obtained from the membership probability analysis, especially at bright magnitudes. This might be a fortuitous result. While it is possible that with the proper motion some objects with highly discrepant values are excluded, our data do not allow us to obtain sufficiently accurate precisions for individual stars to perform CMD cleaning;  the $\sigma$ is large.

\section{Conclusions}
\label{conclusions}
We have used multiple $K_\mathrm{s}$-band observations from the VMC survey spanning $\sim1$ yr to measure the proper motions of the different stellar populations in the SMC, the 47 Tuc Galactic globular cluster, and in the MW foreground. We obtained absolute proper motions calibrated with respect to background galaxies and we analysed separately individual VISTA detectors to reduce the influence of systematic uncertainties. The uncertainty associated with the astrometric reference frame is $0.18$ mas yr$^{-1}$.

We identified stellar populations of different types from their location in CMDs at different distances from the cluster centre and combined their proper motions a posteriori to obtain median values. We measured the proper motions of about $172000$ objects, of which $\sim50\%$ are unique sources and of which about $46000$ refer to the SMC, $86000$ to the 47 Tuc cluster in the region at $10^\prime-60^\prime$ from its centre, $10000$ to the MW, and $30000$ to the background galaxy field. The uncertainties of the proper motions for groups of at least $1000$ objects of a similar type range  from $0.04$ to $0.43$ mas yr$^{-1}$.  We find that the proper motions of red clump and RGB stars of the SMC differ by $\sim0.4$ mas yr$^{-1}$ from the proper motion of mostly main sequence stars as derived by HST. This is partly explained by viewing perspective effects and residual systematic uncertainties, but may also suggest that not only the geometry but also the kinematics of the intermediate-age and old stars across the SMC are different.  The SMC is experiencing a strong tidal effect from its interaction with the LMC and this may have a strong influence on the motion of stars associated to this process. A comprehensive study of the proper motion across the entire SMC combined with advanced modelling of its geometry represents a crucial step in our understanding of the kinematics and dynamics of the system. The internal kinematics of the 47 Tuc cluster, using the large number of detected VMC stars in its outer region with respect to the study by \cite{2006ApJS..166..249M} of the cluster core, is the subject of a separate investigation.

The median proper motions in $(\mu_\alpha\cos(\delta)$, $\mu_\delta)$ of the three main stellar components in the tile are: $(+1.16\pm0.07$,$-0.81\pm0.07)$ mas yr$^{-1}$ for the SMC, $(+7.26\pm0.03$,$-1.25\pm0.03)$ mas yr$^{-1}$ for 47 Tuc, and $(+10.22\pm0.14$,$-1.27\pm0.12)$ mas yr$^{-1}$ for the MW.  Previous determinations of the SMC and 47 Tuc centre of mass proper motions agree with our measurements in the $\mu_\delta$ direction, but differ by $\sim1$ mas yr$^{-1}$ in the $\mu_\alpha\cos(\delta)$ direction. The relative proper motion of 47 Tuc with respect to the SMC agrees with previous studies. Our value of the SMC proper motion represents the first measurements this far out to the West of the galaxy. We have also derived a first estimate of the proper motion of the SMC globular cluster NGC 121, $(\mu_\alpha\cos(\delta)$, $\mu_\delta)=(+2.75\pm0.26$,$-4.67\pm0.47)$ mas yr$^{-1}$.

\begin{acknowledgements}
We thank Ralf-Dieter Scholz for constructive conversations during the preparation stage and the anonymous referee for indicating revisions that have improved our analysis and results. We also thank the Cambridge Astronomy Survey Unit (CASU) and the Wide Field Astronomy Unit (WFAU) in Edinburgh for providing calibrated data products under the support of the Science and Technology Facility Council (STFC) in the UK. RdG acknowledges research support from the National Natural Science Foundation of China (grant 11373010). This research has made use of the NASA/IPAC Extragalactic Database (NED) which is operated by the Jet Propulsion Laboratory, California Institute of Technology, under contract with the National Aeronautics and Space Administration.
\end{acknowledgements}

\bibliographystyle{aa}
\bibliography{my.bib}

\begin{appendix}

\section{Population boxes}
Tables \ref{tucregions} and \ref{mixedregions} indicate the precise boundaries of the regions used in this study to select stars in 47 Tuc, the SMC, and the MW, as well as the region populated by background galaxies. These are the regions shown in Fig.~\ref{pmcmd} (middle and right panels), while boundaries for boxes shown in Fig.~\ref{pmcmd} (left panel) are defined as in \citet{2014A&A...562A..32C} but adjusted to the distance, metallicity, and extinction of the SMC as explained in Sect.~\ref{pmsec}. In particular, in both tables Col. $1$ lists the CMD region and Cols. $2-5$ list the equations for the different boundaries.

\begin{table*}
\caption{Stellar regions in Fig.~\ref{pmcmd}}
\small
\label{tucregions}
\[
\begin{array}{lllllr}
\hline \hline
\mathrm{Region} & \multicolumn{4}{c}{\mathrm{Boundaries}} & \\
\hline
\mathrm{E}_1 & 18.2 < K_s < 19.6 & K_s<-30\times(J-K_s)+36.4 & K_s >-15\times(J-K_s)+25 &  \\
\mathrm{F} & (J-K_s)<0.86 & K_s>-8.75\times(J-K_s)+22.37 & K_s<4.7\times(J-K_s)+13.49 & \\
\mathrm{J} & (J-K_s)>0.36 & K_s>-9.167\times(J-K_s)+21.15 & K_s<-2.5\times(J-K_s)+19 & K_s<-15\times(J-K_s)+25 \\
\mathrm{K} & 13.1<K_s<16.6 & K_s>-15\times(J-K_s)+25 & K_s<-8.75\times(J-K_s)+22.37 \\
\mathrm{L} & 13.1<K_s<19.6 & 0.86<(J-K_s)<2.16 & K_s>-8.75\times(J-K_s)+22.37 & K_s<4.7\times(J-K_s)+13.49\\
\mathrm{M} & 11.8<K_s<17.85 & 0.2<(J-K_s)<0.81 & K_s<-9.167\times(J-K_s)+21.15 & K_s<-15\times(J-K_s)+25 \\
\mathrm{N} & 16.6<K_s<18.2 & K_s<-30\times(J-K_s)+36.4& K_s>-15\times(J-K_s)+25& \\
\mathrm{O} & K_s<19.6 & K_s>4.7\times(J-K_s)+13.49 & K_s>-30\times(J-K_s)+36.4 & & \\
\noalign{\smallskip}

\hline
\hline
\end{array}
\]
\end{table*}

\begin{table*}
\caption{Stellar regions in Fig.~\ref{pmcmd}}
\small
\label{mixedregions}
\[
\begin{array}{lllllr}
\hline \hline
\mathrm{Region} & \multicolumn{4}{c}{\mathrm{Boundaries}} & \\
\hline
\mathrm{E}_1 & K_s < 19.6 & K_s<-30\times(J-K_s)+36.4 & K_s >-15\times(J-K_s)+25 &  K_s>3.6\times(J-K_s)+16.03\\
\mathrm{E}_2 & K_s>16.6 & K_s<-30\times(J-K_s)+36.4 & K_s>-15\times(J-K_s)+25 & K_s<3.6\times(J-K_s)+15.33 \\
\mathrm{F} & (J-K_s)<0.86 & K_s>-30\times(J-K_s)+36.4  & K_s>-8.75\times(J-K_s)+22.37  & K_s<3.6\times(J-K_s)+15.33 \\
\mathrm{J} & (J-K_s)>0.36 & K_s>-9.167\times(J-K_s)+21.15 & K_s<-2.5\times(J-K_s)+19 & K_s<-15\times(J-K_s)+25 \\
\mathrm{K} & 13.1<K_s<16.6 & K_s>-15\times(J-K_s)+25 & K_s<-8.75\times(J-K_s)+22.37 \\
\mathrm{L} & 13.1<K_s<19.6 & 0.86<(J-K_s)<2.16 & K_s>-8.75\times(J-K_s)+22.37 & \\
\mathrm{M} & 11.8<K_s<17.85 & 0.2<(J-K_s)<0.81 & K_s<-9.167\times(J-K_s)+21.15 & K_s<-15\times(J-K_s)+25 \\
\mathrm{N} & K_s<3.6\times(J-K_s)+16.03 & K_s<-30\times(J-K_s)+36.4 & K_s>-15\times(J-K_s)+25 & K_s>3.6\times(J-K_s)+15.33 \\
\mathrm{O} & K_s<19.6 & (J-K_s)<0.86 & K_s>3.6\times(J-K_s)+15.33 & K_s>-30\times(J-K_s)+36.4 & \\
\noalign{\smallskip}

\hline
\hline
\end{array}
\]
\end{table*}

\section{Proper motion and trends}
\label{pmtrends}

\begin{figure*}
\resizebox{\hsize}{!}{\includegraphics{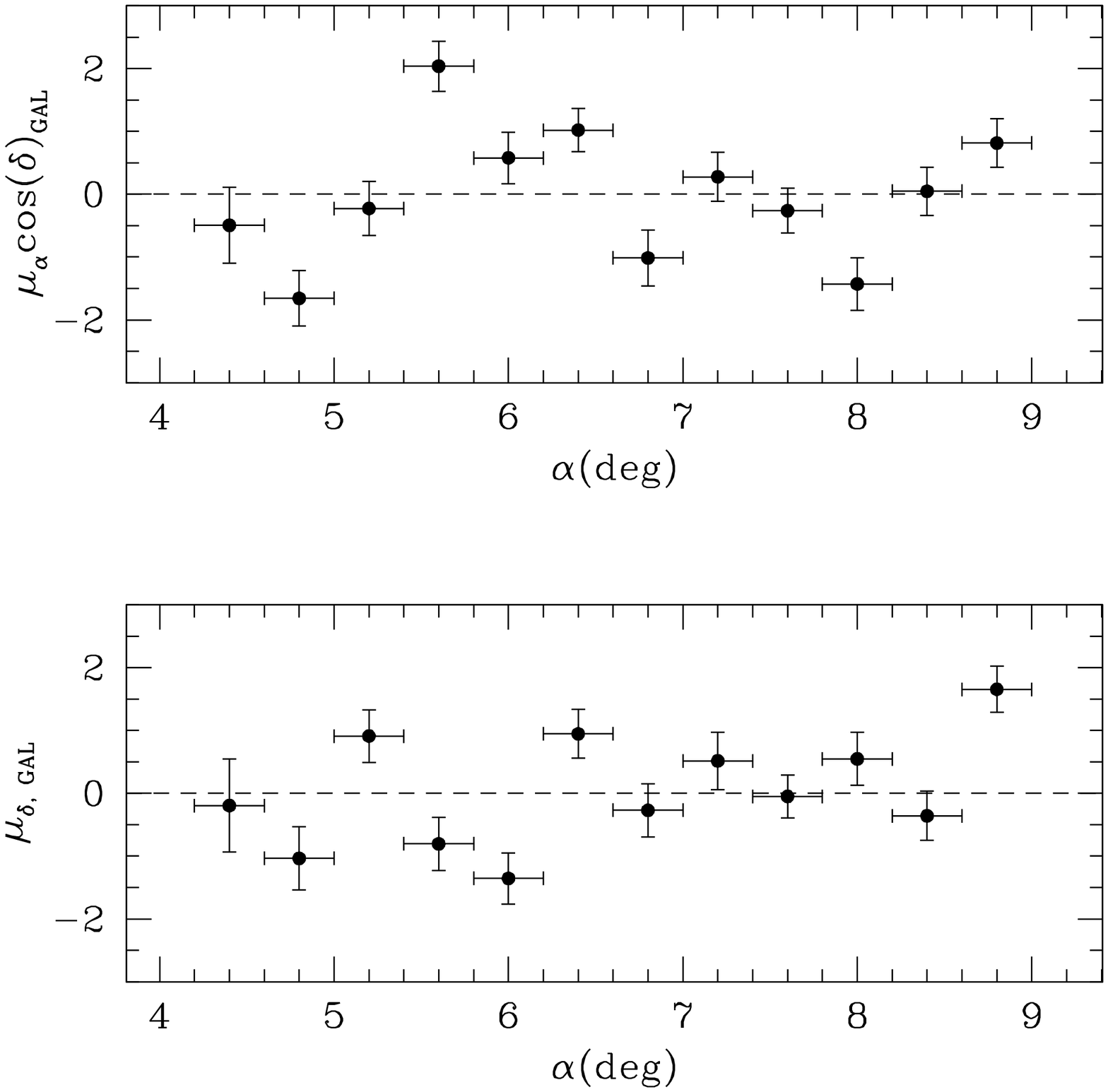}
\includegraphics{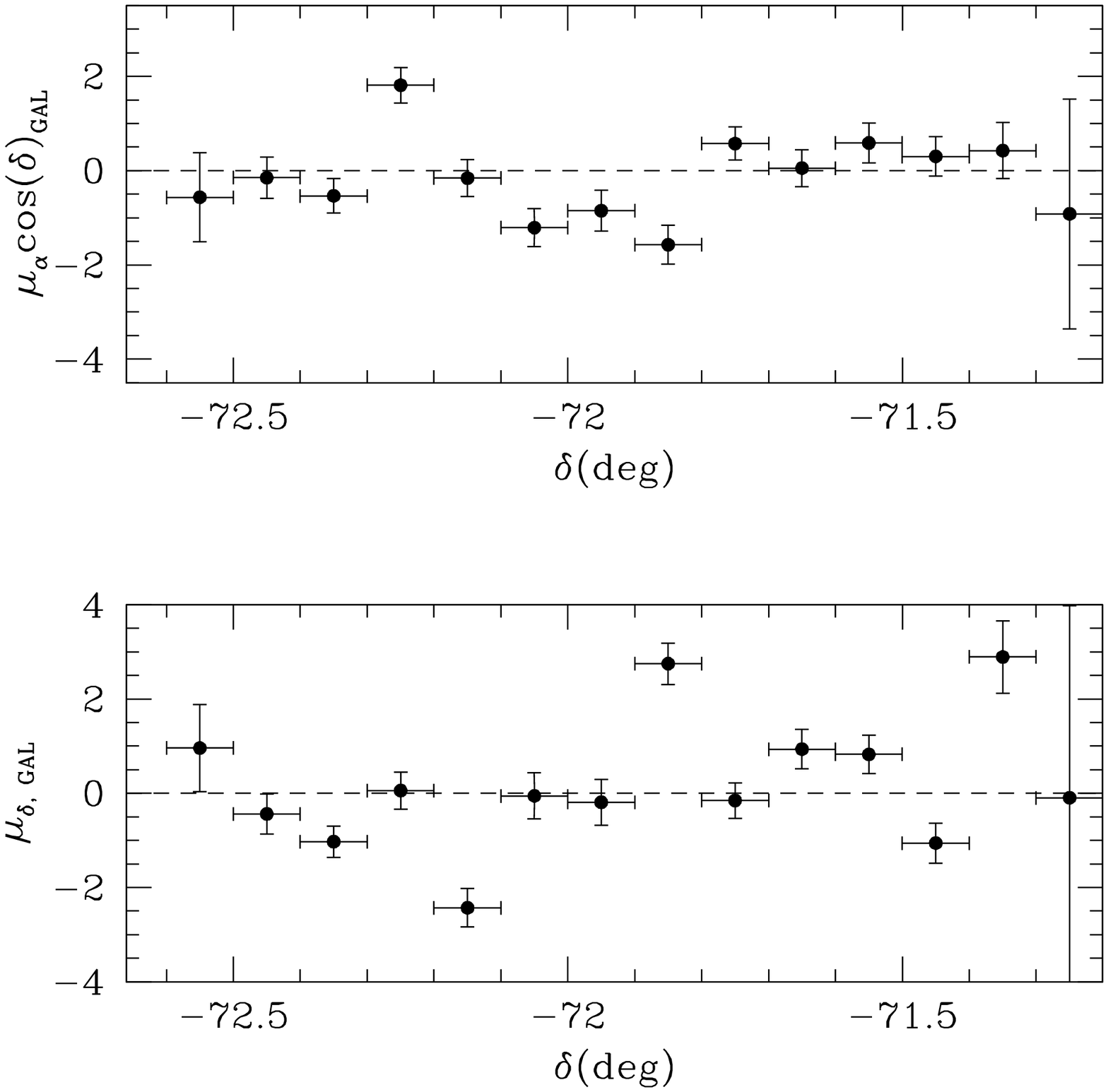}}
\caption{Proper motions in mas yr$^{-1}$ for the $\mu_\alpha\cos(\delta)$ component (top) and the $\mu_\delta$ component (bottom) as a function of right ascension ($\alpha$; left) and declination ($\delta$; right). Lines indicate the medians of all measurements and their uncertainties. Each point corresponds to the median of $1000-5000$ galaxies. Error bars correspond to the bin size in deg ($0.04$ deg and $ $ deg for the $\mu_\alpha\cos(\delta)$ and the $\mu_\delta$ components, respectively) used to derive median proper motions (horizontal) and to the error in the medians (vertical).}
\label{galradec}
\end{figure*}

Figures \ref{galradec}, instead, shows the proper motions as a function of positional coordinates for the sample of background galaxies.  These figures are discussed in Sect.~\ref{systematics}.

\begin{figure*}
\resizebox{\hsize}{!}{\includegraphics{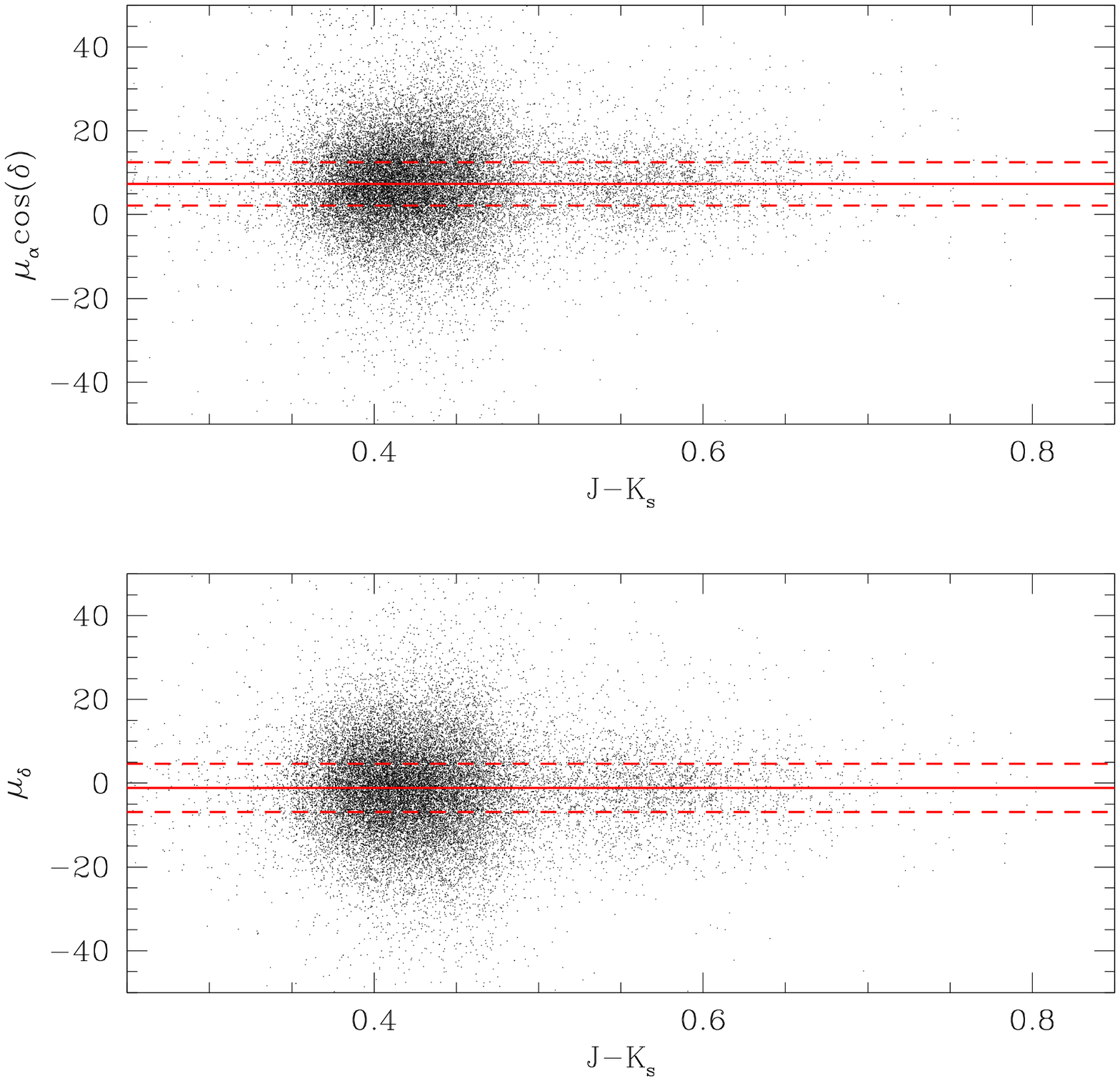}
\includegraphics{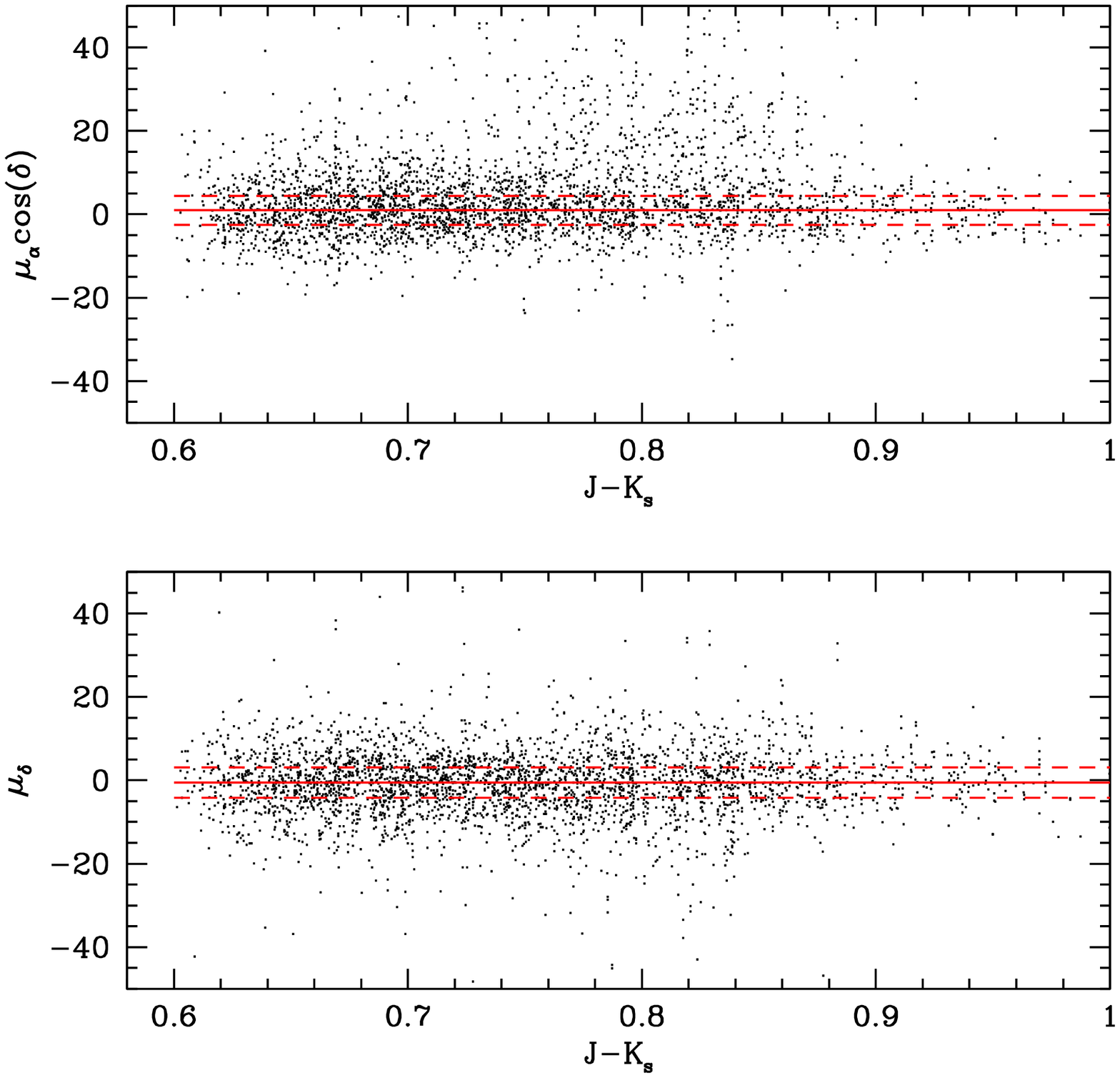}
\includegraphics{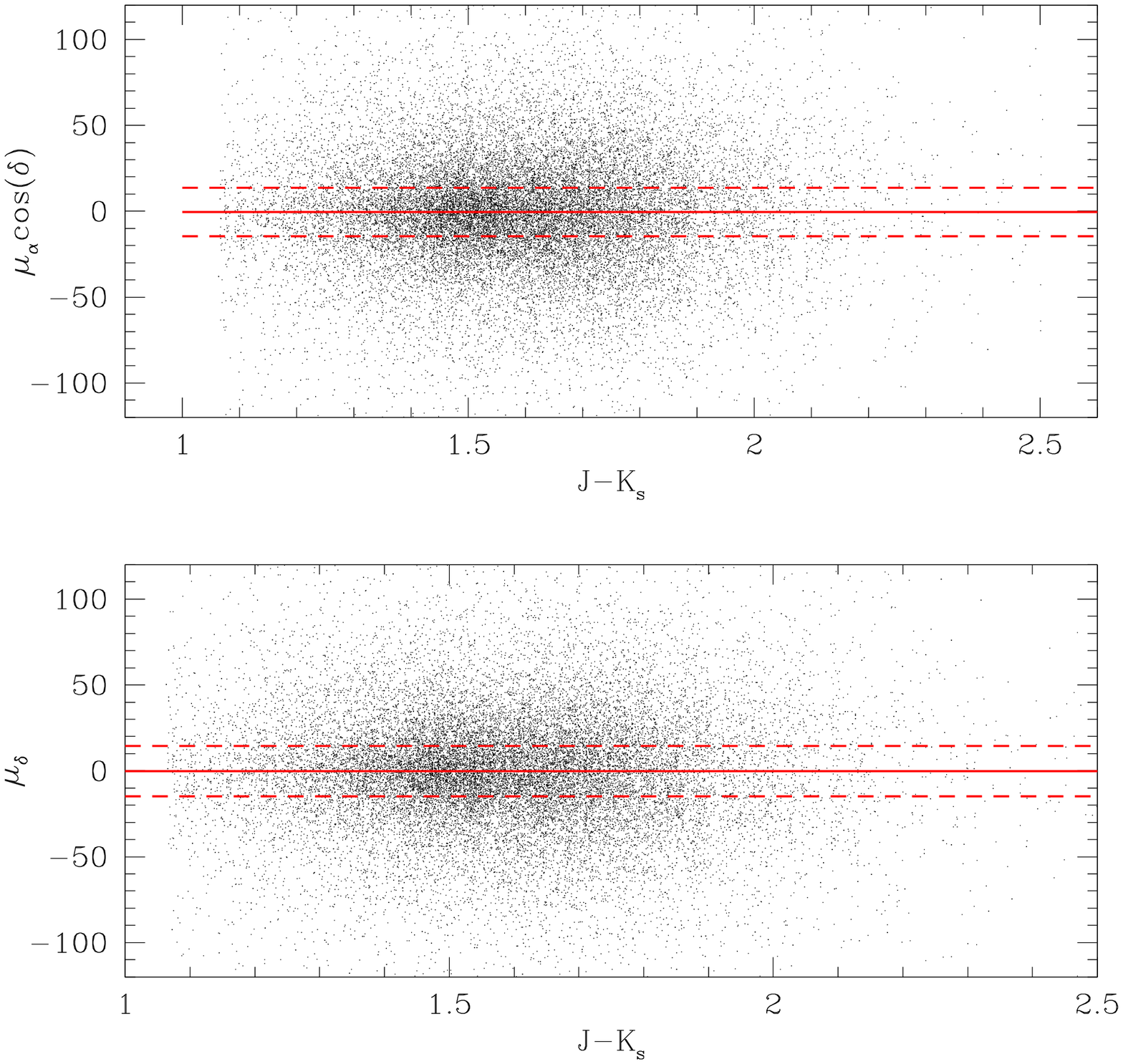}}
\caption{Proper motion in mas yr$^{-1}$ of three different types of objects as a function of $(J-K_\mathrm{s})$ colour. (left) 47 Tuc upper main sequence, subgiant, and RGB stars, those in region M of Fig.~\ref{pmcmd}-middle. (middle) SMC RGB stars, those in region K of Fig.~\ref{pmcmd}-left. (right) Background galaxies. The continuous lines indicate the median proper motion values for each group of objects and dashed lines mark the $1\sigma$ dispersion.}
\label{colfigs}
\end{figure*}

\begin{figure*}
\resizebox{\hsize}{!}{\includegraphics{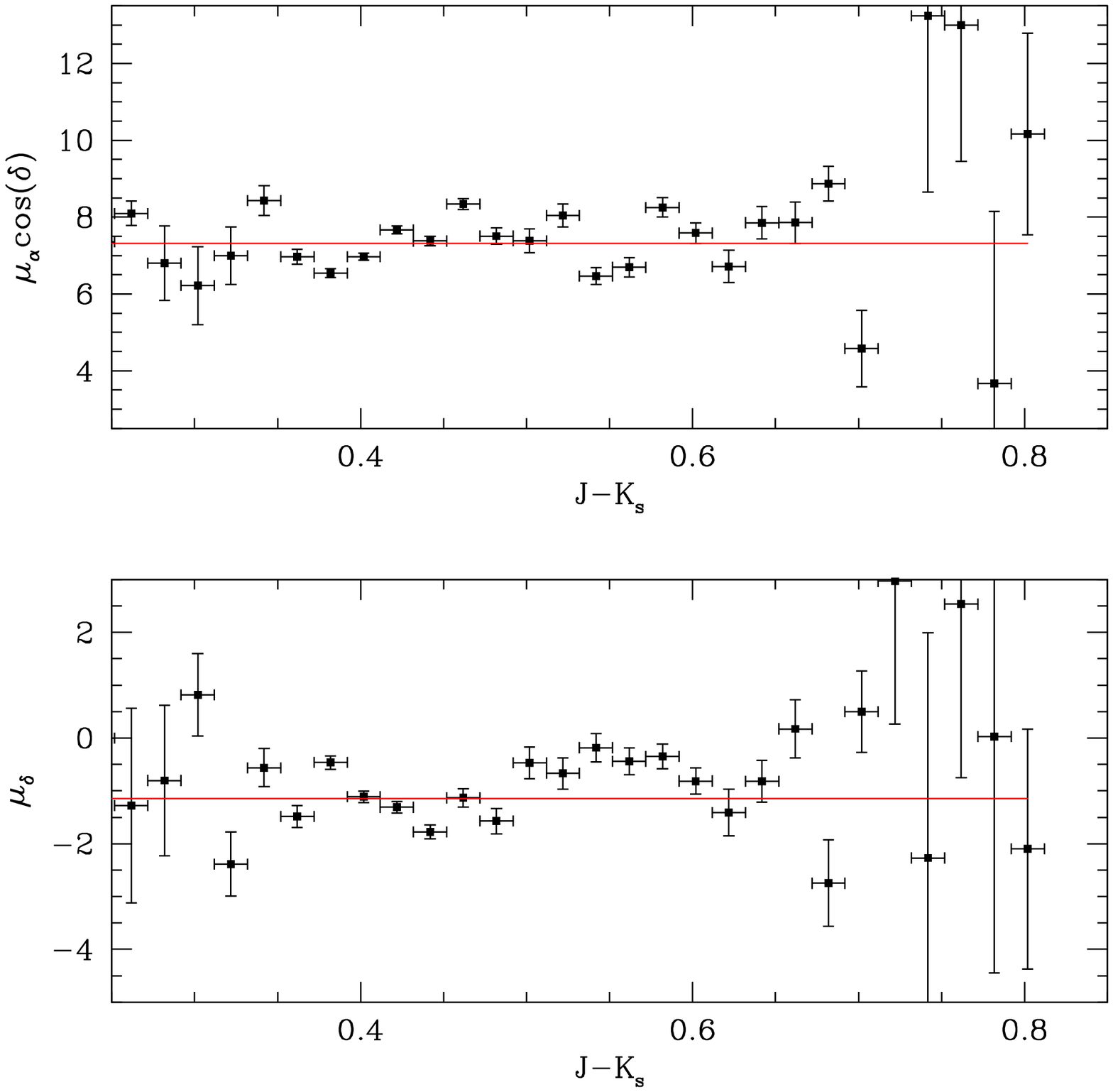}
\includegraphics{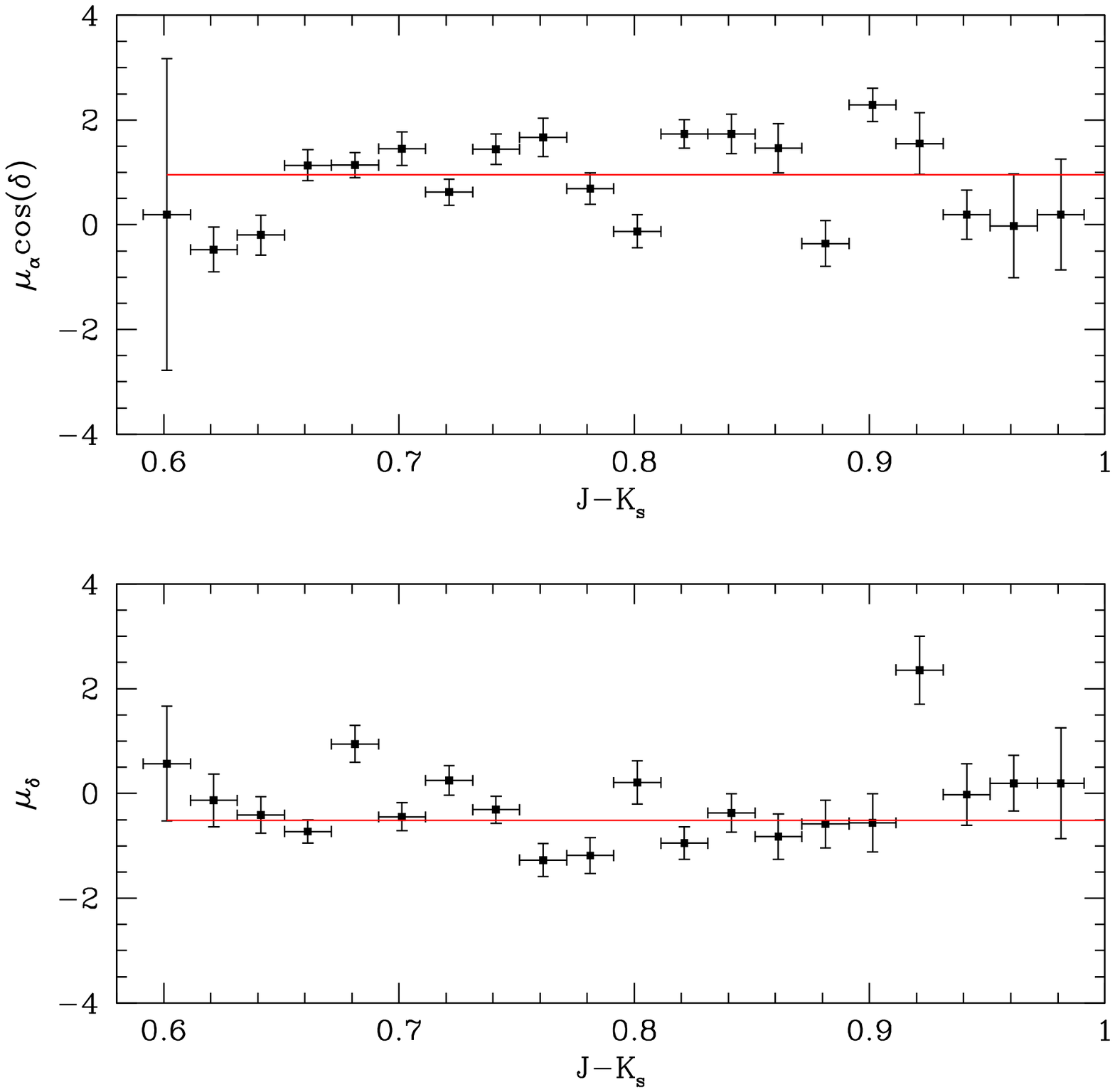}
\includegraphics{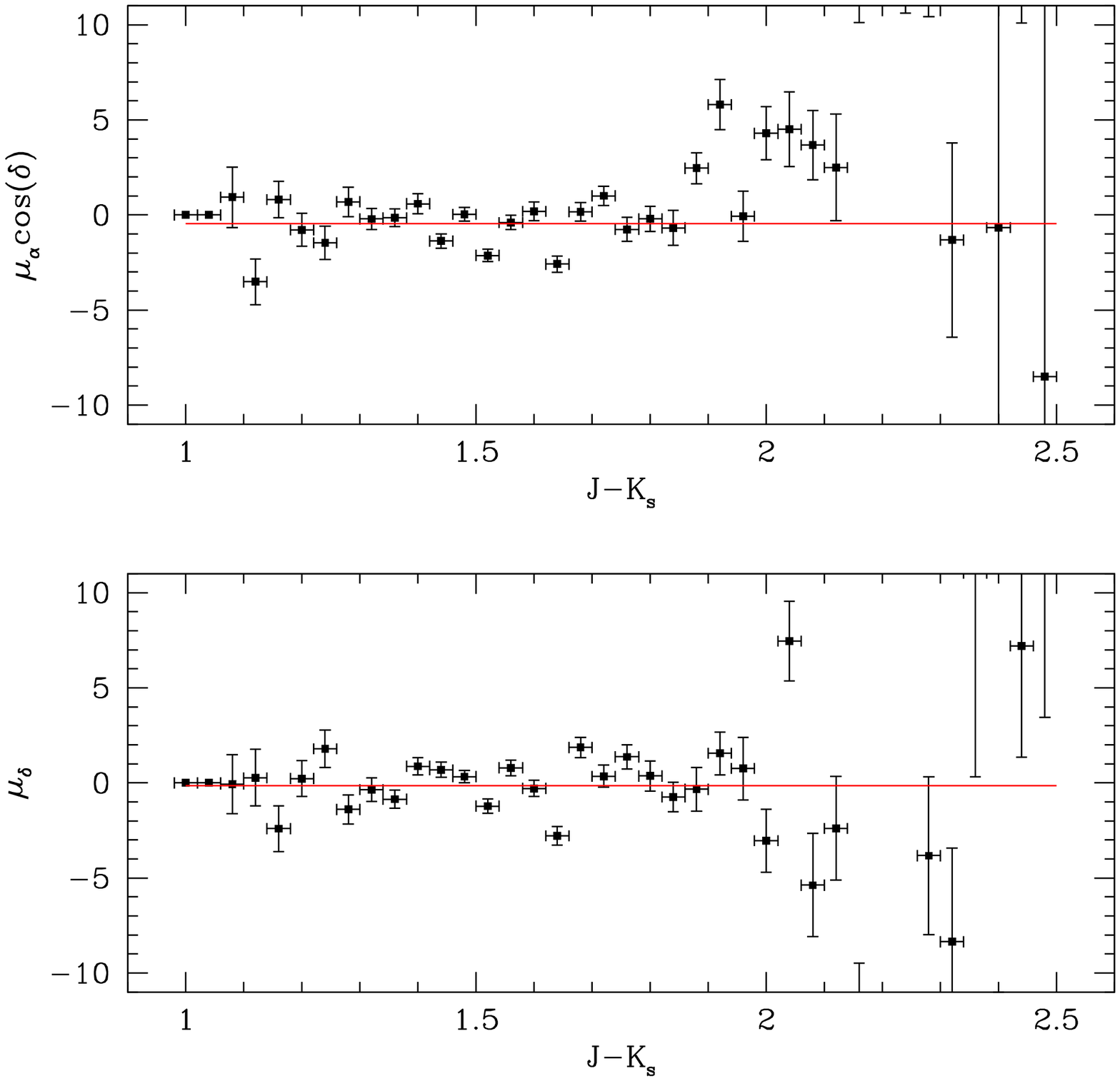}}
\caption{As Fig.~\ref{colfigs} but binned in $(J-K_\mathrm{s})$ colour where the bin size is $0.02$ mag in the left and right panels and $0.04$ mag in the middle panel. The bluest bins in the right panel have very few data points (cfg.~Fig.~\ref{colfigs}).}
\label{colfigsbin}
\end{figure*}

Figure \ref{colfigs} shows the proper motion as a function of colour for some types of objects that together span the colour range of all sources considered in this study. They indicate that no significant trend is present within the $1\sigma$ uncertainty associated with the mean values indicating that possible differential atmospheric refraction effects are small. To check this further we have created a similar figure (Fig.~\ref{colfigsbin}) by binning the proper motions in colour. We noticed that apart from where the number of sources is small, at the edges of the colour ranges, there is no significant trend for either of the groups of 47 Tuc stars, SMC RGB stars, and background galaxies. 
The higher proper motions found at $(J-K_\mathrm{s})=0.7-0.9$ mag (Fig.~\ref{colfigs} top-middle panel) correspond to MW stars (cfg. Fig.~\ref{pmcmd}) that, however, have no effect on the medial value of the proper motions for the region.

\end{appendix}

\end{document}